\documentclass[aps,prd,twocolumn,nofootinbib,superscriptaddress,longbibliography]{revtex4-2}

\usepackage[T1]{fontenc}
\usepackage[utf8]{inputenc}
\usepackage{lmodern}
\usepackage{amsmath,amssymb,amsthm,mathtools}
\usepackage{graphicx}
\usepackage{xcolor}
\usepackage{hyperref}
\usepackage{physics}
\usepackage{bm}
\usepackage{booktabs}
\usepackage{multirow}
\usepackage{slashed}
\usepackage{placeins}
\usepackage{float}
\hypersetup{
	colorlinks=true,
	linkcolor=blue,
	citecolor=blue,
	urlcolor=blue
}

\begin{document}
	\title{Null Tests and Lepton Universality in $\Xi_{cc}$ Baryon Decays}
	
	\author{Hindi Zouhair}
	\email{hindizouhair@gmail.com}
	\affiliation{Laboratory of High Energy Physics (LHEP-MS), Mohammed V University, Rabat, Morocco}
	
	\date{\today}
	\begin{abstract}
		We develop a precision framework for doubly charmed baryon decays based
		on symmetry-protected observables and effective-field-theory diagnostics.
		In nonleptonic $\Xi_{cc}$ decays, we construct a null combination of
		widths that vanishes in the heavy-diquark factorization limit, providing
		a direct probe of nonfactorizable QCD dynamics.
		For semileptonic decays, we identify the light-lepton universality ratio
		$R_{\Xi_c}^{\mu e}$ as an observable in which leading hadronic
		normalization cancels at the amplitude level, yielding direct sensitivity
		to short-distance charged-current interactions. Percent-level precision
		probes $|C_{V_L}^{\mu}|\sim \mathcal{O}(10^{-2})$, whereas
		$\mathcal{O}(10^{-1})$ deformations induce order-one deviations. Scalar
		contributions remain parametrically suppressed.
		Combining baryonic and mesonic inputs, we show that $\Xi_{cc}$ decays
		constrain the same short-distance interaction with complementary scaling,
		lifting degeneracies inherent to meson-only analyses. Mapping to a
		charged-vector benchmark demonstrates sensitivity to multi-TeV
		new-physics scales. These results establish doubly charmed baryons as an
		independent probe of charged-current interactions beyond the Standard
		Model.
	\end{abstract}
	\maketitle
\section{Introduction}
Hadrons containing heavy quarks provide a powerful laboratory for studying the interplay between strong and weak interactions. In systems with a single heavy quark, symmetry-based approaches and expansions in $\Lambda_{\rm QCD}/m_Q$ have enabled the construction of precision observables with controlled theoretical uncertainties, allowing for a systematic separation between hadronic effects and short-distance dynamics~\cite{Isgur:1989vq,Neubert:1993mb,Manohar:2000dt}. Extending such precision analyses to new hadronic systems is therefore a natural step toward broadening the scope of flavor physics. The observation of the doubly charmed baryon $\Xi_{cc}^{++}$ by the
LHCb Collaboration~\cite{LHCb:2017iph}, followed by the measurement of additional decay modes~\cite{LHCb:2018pcs}, opens access to a distinct sector of QCD in which two heavy quarks form a compact diquark configuration interacting with a light spectator quark. This system combines the predictive structure of heavy-quark symmetry with nontrivial baryonic dynamics associated with the light degrees of freedom. The heavy $(cc)$ subsystem admits an expansion in $\Lambda_{\rm QCD}/m_c$, while the baryonic structure introduces correlations that are absent in mesonic systems ~\cite{Savage:1990di,Neubert:1993mb,Manohar:2000dt}. As a result, doubly charmed baryons probe QCD dynamics and charged-current interactions in a regime complementary to heavy-meson systems. Existing studies of doubly heavy baryons have focused primarily on spectroscopy, lifetimes, and model-dependent predictions for exclusive
decay rates~\cite{Kiselev:2001fw,Karliner:2017qjm,Bigi:2017jbd,Cheng:2018hwl,Shi:2019hbf,Ebert:2004ck}. While these analyses have established the basic phenomenology of the $\Xi_{cc}$ system, a precision framework based on observables with systematically controlled uncertainties and a direct connection to short-distance dynamics is still absent. In particular, a systematic construction of observables that satisfy symmetry-protected relations and admit a direct interpretation in terms of effective operators remains to be developed. In this work, we develop such a framework. Our approach is to exploit the heavy-diquark structure of $\Xi_{cc}$ baryons to construct observables with enhanced theoretical control. In nonleptonic decays, we define null combinations of decay widths that vanish in the factorization and heavy-diquark limits, thereby isolating nonfactorizable QCD contributions in a model-independent way~\cite{Beneke:1999br,Buchalla:1995vs}. In semileptonic decays, we focus on the light-lepton universality ratio
\[
R_{\Xi_c}^{\mu e} = \frac{\Gamma(\Xi_{cc} \to \Xi_c \mu \nu)}{\Gamma(\Xi_{cc} \to \Xi_c e \nu)},
\]
which benefits from a strong cancellation of hadronic normalization and provides making it a theoretically controlled null observable whose leading nonvanishing contribution directly isolates subleading QCD dynamics of charged-current interactions. Within a low-energy effective-field-theory (EFT) framework~\cite{Buchmuller:1985jz,Grzadkowski:2010es,Cirigliano:2009wk}, deviations from the Standard Model can be parameterized in terms of Wilson coefficients multiplying dimension-six operators. We show that $R_{\Xi_c}^{\mu e}$ is insensitive to universal rescalings of the charged current, while exhibiting enhanced sensitivity to lepton-nonuniversal vector interactions. Expanding around the Standard Model point,
\[
R_{\Xi_c}^{\mu e} \simeq R_{\Xi_c}^{\mu e}\big|_{\rm SM}\left(1 + 2\,C_{V_L}^{\mu}\right),
\]
so that percent-level measurements directly probe Wilson coefficients at the level $|C_{V_L}^{\mu}| \sim \mathcal{O}(10^{-2})$.
The key result of this work is that doubly charmed baryon decays admit observables that are both theoretically controlled and directly sensitive to short-distance interactions. A central question is whether such observables can probe new physics at a level competitive with existing mesonic constraints and collider searches. Addressing this question is
one of the main goals of the present analysis. The paper is organized as follows. In Sec.~\ref{sec:framework}, we introduce the heavy-diquark framework and construct the nonleptonic null observable $\Delta_{\pi K}$. In Sec.~\ref{sec:semileptonic}, we define
the light-lepton universality ratio $R_{\Xi_c}^{\mu e}$ and derive its Standard Model benchmark. In Sec.~\ref{sec:eft}, we analyze the response of this observable to vector and scalar effective operators. In Sec.~\ref{sec:constraints}, we derive projected constraints on
lepton-nonuniversal charged-current interactions and compare them with
existing light-lepton meson data. In Sec.~\ref{sec:UV_Wprime}, we map the EFT constraints onto a charged-vector benchmark. Section~\ref{sec:conclusion} summarizes our results.
\section{Heavy-diquark framework and construction of null observables}
\label{sec:framework}
A convenient starting point for describing doubly charmed baryons is the heavy-diquark picture, in which the $\Xi_{cc}$ state is treated schematically as
\[
\Xi_{cc} \sim (cc)_{\bar{\mathbf 3}} + q,
\]
where the two charm quarks form a compact color-antitriplet diquark and the light quark $q$ carries the long-distance dynamics. In this description, the $(cc)$ subsystem acts as an approximately pointlike color source interacting with soft gluonic fields. This provides a natural extension of heavy-quark symmetry concepts to baryonic systems and can be formalized within effective field theory descriptions of doubly heavy hadrons~\cite{Savage:1990di,Fleming:2005pd,Neubert:1993mb,Manohar:2000dt}.
For momentum transfers of order $\Lambda_{\rm QCD}$, the diquark can be treated as approximately compact, allowing for a systematic expansion in inverse powers of the heavy scale. The relevant parameter is
\[
\epsilon_{\rm HQ} \sim \frac{\Lambda_{\rm QCD}}{m_c},
\]
which controls deviations from the heavy-diquark limit. At leading order, the $(cc)$ subsystem behaves as a static color source, and decay amplitudes factorize into a short-distance weak transition and a long-distance baryonic matrix element.
Corrections to this limit arise from finite-size and recoil effects of the diquark, as well as from spin-dependent interactions induced by the chromomagnetic operator in the heavy-quark expansion, and scale parametrically as
\[
\delta_{\rm HQ} \sim \mathcal{O}\!\left(\frac{\Lambda_{\rm QCD}}{m_c}\right).
\]
A second source of theoretical uncertainty is provided by flavor-$SU(3)$ breaking,
\[
\delta_{SU(3)} \sim \mathcal{O}\!\left(\frac{m_s - m_{u,d}}{\Lambda_{\rm QCD}}\right),
\]
which becomes relevant when comparing decay channels with different light-quark content.
In addition, nonfactorizable QCD effects, including weak-exchange
topologies, final-state rescattering, and soft-gluon exchange between
the mesonic and baryonic subsystems, generate corrections that are not
captured by the factorized amplitude. We denote these effects collectively
by $\delta_{\rm nonfact}$. Unlike finite-size or recoil corrections,
$\delta_{\rm nonfact}$ is not necessarily suppressed by the heavy-diquark
expansion parameter and can therefore compete with
$\mathcal{O}(\Lambda_{\rm QCD}/m_c)$ corrections. It parameterizes
long-distance QCD dynamics beyond the factorization approximation
~\cite{Beneke:1999br,Buchalla:1995vs,Eakins:2012bb}.
The decay amplitude can therefore be organized schematically as
\begin{equation}
\mathcal{A} = \mathcal{A}_{\rm fact}
\left[
1
+ \mathcal{O}\!\left(\frac{\Lambda_{\rm QCD}}{m_c}\right)
+ \mathcal{O}\!\left(\delta_{SU(3)}\right)
+ \mathcal{O}\!\left(\delta_{\rm nonfact}\right)
\right],
\label{eq:ampExpansion}
\end{equation}
In contrast to heavy-meson systems, where factorization can be justified
in the heavy-quark limit, the baryonic environment admits additional
topologies that are not power suppressed, making null observables a particularly useful tool for isolating nonfactorizable QCD contributions.
where $\mathcal{A}_{\rm fact}$ denotes the leading contribution in the heavy-diquark factorization limit.
This hierarchical structure suggests that observables insensitive to the leading contribution $\mathcal{A}_{\rm fact}$ can be systematically constructed. The central implication is that one can define observables in which the leading contribution cancels identically. In particular, appropriately normalized ratios and linear combinations of decay widths can be constructed such that they vanish in the heavy-diquark factorization limit. Deviations from zero therefore provide a direct measure of subleading effects, including nonfactorizable QCD dynamics and symmetry-breaking corrections. This construction forms the basis of the null tests introduced in the following section.
At the same time, in semileptonic observables such as $R_{\Xi_c}^{\mu e}$, the cancellation of leading hadronic normalization implies that residual sensitivity is dominated by short-distance interactions. This provides a direct link between the heavy-diquark expansion and the effective-field-theory interpretation developed below.
This interplay between symmetry protection, controlled power counting, and cancellation of leading hadronic contributions is the key mechanism that promotes doubly charmed baryons from a purely spectroscopic system to a precision probe of QCD dynamics and charged-current interactions.
\subsection{Benchmark null observable}
\label{sec:nulltest}
In this work, a null test denotes an observable that vanishes in the
heavy-diquark factorization limit. Any nonzero value therefore directly
probes subleading QCD dynamics and symmetry-breaking effects.
To make this construction explicit, we consider the two benchmark channels
\[
\Xi_{cc}^{++}\to \Xi_c^+ \pi^+,
\qquad
\Xi_{cc}^{++}\to \Xi_c^+ K^+ .
\]
The related two-body nonleptonic channels have been analyzed in quark models and pole-model approaches~\cite{Zeng:2022eoc}.
Nonleptonic heavy-hadron decays can be described within the effective
weak Hamiltonian framework~\cite{Buchalla:1995vs,Korner:1992wi}, in which short-distance QCD effects are encoded in Wilson coefficients multiplying four-fermion operators.
In the factorization approximation~\cite{Bauer:1986bm}, the decay amplitude for
$\Xi_{cc}\to \mathcal{B}_c M$
can be written as
\begin{align}
\mathcal A(\Xi_{cc}\to \mathcal{B}_c M)
&=
\frac{G_F}{\sqrt2}\,
V_{cq_1}V_{uq_2}^*\,
a_1\,f_M\,q^\mu
\nonumber\\
&\quad \times
\langle \mathcal B_c|\bar q\,\gamma_\mu(1-\gamma_5)c|\Xi_{cc}\rangle ,
\label{eq:amp_fact}
\end{align}
where $a_1$ encodes short-distance QCD corrections and the factorized
structure separates the mesonic and baryonic dynamics~\cite{Beneke:1999br,Neubert:1997uc}.
The hadronic matrix element is parameterized in terms of vector and axial-vector form factors,
\begin{align}
\langle \mathcal B_c(p')|\bar q\gamma^\mu c|\Xi_{cc}(p)\rangle
&=
\bar u_{\mathcal B_c}(p')
\Bigg[
f_1(q^2)\gamma^\mu
+i\frac{f_2(q^2)}{m_{\Xi_{cc}}}\sigma^{\mu\nu}q_\nu
\nonumber\\
&\hspace{1.0cm}
+\frac{f_3(q^2)}{m_{\Xi_{cc}}}q^\mu
\Bigg]
u_{\Xi_{cc}}(p),
\label{eq:vectorFF}
\\[1mm]
\langle \mathcal B_c(p')|\bar q\gamma^\mu\gamma_5 c|\Xi_{cc}(p)\rangle
&=
\bar u_{\mathcal B_c}(p')
\Bigg[
g_1(q^2)\gamma^\mu
+i\frac{g_2(q^2)}{m_{\Xi_{cc}}}\sigma^{\mu\nu}q_\nu
\nonumber\\
&\hspace{1.0cm}
+\frac{g_3(q^2)}{m_{\Xi_{cc}}}q^\mu
\Bigg]\gamma_5
u_{\Xi_{cc}}(p).
\label{eq:axialFF}
\end{align}
which follows the standard decomposition of baryonic weak currents~\cite{Cheng:2018hwl,Eakins:2012bb}.
Contracting with the meson momentum $q^\mu$ in Eq.~\eqref{eq:amp_fact}
leads to a simplification,
\[
q_\mu \sigma^{\mu\nu} q_\nu = 0,
\]
so that the contributions proportional to $f_2$ and $g_2$ vanish. Retaining only the leading form factors $f_1$ and $g_1$, the amplitude reduces to
\begin{equation}
\mathcal A
\simeq
\frac{G_F}{\sqrt2}\,
V_{cq_1}V_{uq_2}^*\,
a_1 f_M\,
\bar u_{\mathcal B_c}
\left(
f_1 \slashed q
-
g_1 \slashed q\,\gamma_5
\right)
u_{\Xi_{cc}}.
\label{eq:ampLeading}
\end{equation}
The corresponding spin-averaged squared amplitude takes the form
\begin{equation}
\overline{|\mathcal A|^2}
=
\frac{G_F^2}{2}\,
|V_{cq_1}V_{uq_2}^*|^2\,
a_1^2 f_M^2\,
\mathcal H_M,
\end{equation}
where $\mathcal H_M$ is a kinematic function depending on hadronic
form factors and masses. Its explicit form is not required here,
since it is identical up to higher-order corrections in
$\Lambda_{\rm QCD}/m_c$ and $SU(3)$ breaking, and therefore cancels in
ratios at leading power The two-body decay width is given by
\begin{equation}
\Gamma(\Xi_{cc}\to \mathcal B_c M)
=
\frac{|\vec p\,|}{8\pi m_{\Xi_{cc}}^2}\,
\overline{|\mathcal A|^2},
\end{equation}
with $|\vec p\,|$ determined by standard phase space.
At leading order in the heavy-diquark and factorization limits, the
same hadronic structure enters both benchmark channels. As a result,
the ratio of widths simplifies to
\begin{equation}
\frac{\Gamma(\Xi_{cc}^{++}\to \Xi_c^+ K^+)}
{\Gamma(\Xi_{cc}^{++}\to \Xi_c^+ \pi^+)}
\simeq
\frac{|V_{us}|^2}{|V_{ud}|^2}
\frac{f_K^2}{f_\pi^2}
\frac{|\vec p_K|}{|\vec p_\pi|}.
\label{eq:ratioKpi}
\end{equation}
The cancellation of the baryonic matrix element at leading order is the
central mechanism that enables the null-test construction.
This cancellation of the leading hadronic structure motivates the
definition of the null observable
\begin{equation}
\Delta_{\pi K}
=
\frac{
	\Gamma(\Xi_{cc}^{++}\to \Xi_c^+ K^+)
	-
	\kappa_{\pi K}\,
	\Gamma(\Xi_{cc}^{++}\to \Xi_c^+ \pi^+)
}{
	\Gamma(\Xi_{cc}^{++}\to \Xi_c^+ K^+)
	+
	\kappa_{\pi K}\,
	\Gamma(\Xi_{cc}^{++}\to \Xi_c^+ \pi^+)
},
\label{eq:Deltapik}
\end{equation}
with
\begin{equation}
\kappa_{\pi K}
=
\frac{|V_{us}|^2}{|V_{ud}|^2}
\frac{f_K^2}{f_\pi^2}
\frac{|\vec p_K|}{|\vec p_\pi|}.
\end{equation}
This relation demonstrates that the observable isolates deviations from
the factorized limit, with the leading hadronic contribution canceling
at this order.
Using physical inputs from the Particle Data Group~\cite{PDG:2024}, we obtain
\[
\kappa_{\pi K} \simeq 6.9\times 10^{-2}.
\]
By construction,
\begin{equation}
\Delta_{\pi K}^{\rm fact} = 0,
\end{equation}
so that any nonzero value directly signals deviations from the leading
heavy-diquark factorization limit. The smallness of $\kappa_{\pi K}$
further enhances the sensitivity to channel-dependent corrections.
To illustrate this explicitly, we parameterize deviations through
\[
\Gamma_K = \Gamma_K^{(0)}(1+\epsilon_K),
\qquad
\Gamma_\pi = \Gamma_\pi^{(0)},
\]
which yields
\begin{equation}
\Delta_{\pi K} = \frac{\epsilon_K}{2+\epsilon_K}.
\label{eq:DeltaEpsilon}
\end{equation}
\begin{figure}[t]
	\centering
	\includegraphics[width=\columnwidth]{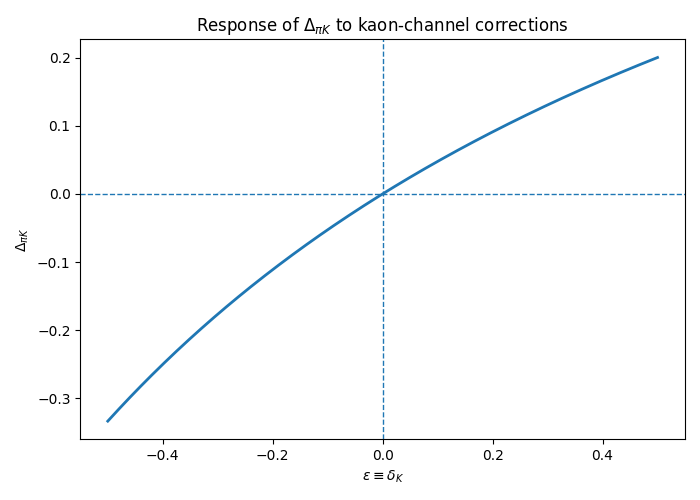}
	\caption{
		Response of the null observable $\Delta_{\pi K}$ to a channel-dependent correction $\epsilon_K$ in the kaon mode. The observable vanishes in the factorized limit and deviates monotonically from zero as subleading effects are introduced, illustrating its sensitivity to nonfactorizable QCD dynamics and $SU(3)$ breaking.
	}\label{fig:DeltaPiKepsilon}
\end{figure}
This expression demonstrates that $\Delta_{\pi K}$ vanishes in the
factorized limit and responds monotonically to corrections, making it
a clean probe of nonfactorizable QCD dynamics and $SU(3)$ breaking.
We expand the decay widths around the heavy-diquark factorization limit as
\begin{equation}
\Gamma_i = \Gamma_i^{(0)}\left[
1 + \mathcal O\!\left(
\frac{\Lambda_{\rm QCD}}{m_c},\,\delta_{SU(3)},\,\delta_{\rm nonfact}
\right)
\right].
\end{equation}
Using $\Delta_{\pi K}^{\rm fact}=0$, it follows that the first
nonvanishing contribution to $\Delta_{\pi K}$ is linear in these
subleading corrections.
\begin{equation}
\Delta_{\pi K}
\sim
\mathcal O\!\left(
\frac{\Lambda_{\rm QCD}}{m_c},\,
\delta_{SU(3)},\,
\delta_{\rm nonfact}
\right).
\end{equation}
Importantly, this sensitivity arises with reduced dependence on
nonperturbative normalization, thereby enhancing the theoretical
robustness of the observable.
\section{Semileptonic decays and LFU observables}
\label{sec:semileptonic}
Semileptonic decays of doubly charmed baryons provide a complementary
probe of short-distance charged-current dynamics, with a theoretical
structure distinct from that of nonleptonic modes. We consider transitions of the form
\begin{equation}
\Xi_{cc} \to \Xi_c \,\ell \,\nu,
\end{equation}
Such semileptonic transitions have been studied in model-based analyses
of doubly charmed baryons~\cite{Geng:2022yxb}.
with $\ell=e,\mu$ in the present analysis. In these processes, the
leptonic current is perturbatively calculable, and suitably constructed
ratios of decay rates significantly reduce the dependence on overall
hadronic normalization. This feature makes semileptonic observables
particularly suitable for precision tests of lepton-flavor universality
(LFU)~\cite{Neubert:1993mb,Manohar:2000dt}.
For the specific transition $\Xi_{cc}\to \Xi_c \ell \nu$, the available
phase space is bounded by
\begin{equation}
q^2_{\max}=(m_{\Xi_{cc}}-m_{\Xi_c})^2 < m_\tau^2,
\end{equation}
so that decays involving $\tau$ leptons are kinematically forbidden.
As a consequence, LFU tests in this channel are restricted to light
leptons. This simplifies the interpretation of LFU ratios, as it removes
the large kinematic hierarchy associated with the $\tau$ mass and
renders the observable sensitive to small light-lepton effects and
possible nonuniversal short-distance contributions.
We define the light-lepton universality ratio
\begin{equation}
R_{\Xi_c}^{\mu e}
=
\frac{\Gamma(\Xi_{cc} \to \Xi_c \mu \nu)}
{\Gamma(\Xi_{cc} \to \Xi_c e \nu)}.
\label{eq:RXi_me}
\end{equation}
At leading order in the heavy-diquark expansion, the same hadronic
matrix elements enter both decay modes, implying that their
normalization largely cancels in the ratio~\cite{Cirigliano:2009wk}.
As a result, $R_{\Xi_c}^{\mu e}$ is primarily sensitive to phase-space
effects and to potential lepton-nonuniversal contributions to the
charged current.
\subsection{Differential decay rate}
The differential decay rate for the semileptonic transition
$\Xi_{cc}\to \Xi_c \ell \nu$ can be written in the standard form
\begin{equation}
\frac{d\Gamma}{dq^2}
=
\frac{G_F^2 |V_{cq}|^2}{192\pi^3 m_{\Xi_{cc}}^3}
\,\lambda^{1/2}(q^2)\,
L_{\mu\nu} H^{\mu\nu},
\label{eq:dGamma_general}
\end{equation}
where $q^2$ denotes the dilepton invariant mass and
\begin{equation}
\lambda(q^2)
=
\lambda(m_{\Xi_{cc}}^2,m_{\Xi_c}^2,q^2)
\end{equation}
is the Källén function. The leptonic tensor $L_{\mu\nu}$ is
perturbatively calculable, while the hadronic tensor $H^{\mu\nu}$
encodes the nonperturbative QCD dynamics of the baryonic transition.
The hadronic tensor is constructed from the vector and axial-vector
currents. In the heavy-diquark limit, the dominant contributions are governed by
the leading form factors $f_1(q^2)$ and $g_1(q^2)$, while subleading
structures are parametrically suppressed by
$\mathcal O(\Lambda_{\rm QCD}/m_c)$~\cite{Savage:1990di,Fleming:2005pd}.
After contraction of leptonic and hadronic tensors, the differential
rate can be expressed schematically as
\begin{equation}
\frac{d\Gamma}{dq^2}
\propto
\lambda^{1/2}(q^2)
\left[
(f_1^2+g_1^2)\,F_V(q^2,m_\ell^2)
+
(f_1^2-g_1^2)\,F_A(q^2,m_\ell^2)
\right],
\end{equation}
where the functions $F_V$ and $F_A$ encode the kinematic dependence on
$q^2$ and the lepton mass. In the Standard Model, the difference between
electron and muon channels arises entirely from the finite lepton-mass
dependence entering these kinematic functions.
The LFU ratio is obtained by integrating over phase space,
\begin{equation}
R_{\Xi_c}^{\mu e}
=
\frac{
	\displaystyle
	\int_{m_\mu^2}^{(m_{\Xi_{cc}}-m_{\Xi_c})^2}
	dq^2\,\frac{d\Gamma(\mu)}{dq^2}
}{
	\displaystyle
	\int_{m_e^2}^{(m_{\Xi_{cc}}-m_{\Xi_c})^2}
	dq^2\,\frac{d\Gamma(e)}{dq^2}
}.
\label{eq:RXi_me_integral}
\end{equation}
For a benchmark estimate, we adopt a dipole parametrization of the
leading form factors,
\begin{equation}
f_1(q^2)=\frac{f_1(0)}{(1-q^2/m_V^2)^2},
\qquad
g_1(q^2)=\frac{g_1(0)}{(1-q^2/m_A^2)^2},
\end{equation}
with representative values $f_1(0)=0.60$, $g_1(0)=0.55$, and pole masses
$m_V=2.1~\mathrm{GeV}$ and $m_A=2.5~\mathrm{GeV}$. These values are
consistent with typical quark-model and QCD-based estimates of
heavy-baryon form factors~\cite{Ebert:2004ck,Faustov:2016pal,Cheng:2018hwl,Detmold:2015aaa,Meinel:2016dqj}.
A numerical evaluation of the phase-space integrals yields
\begin{align}
\Gamma(\Xi_{cc} \to \Xi_c e \nu)
&\simeq 2.02\times 10^{-15},
\nonumber\\
\Gamma(\Xi_{cc} \to \Xi_c \mu \nu)
&\simeq 1.97\times 10^{-15}.
\label{eq:Gamma_mue_num}
\end{align}
corresponding to the Standard Model benchmark
\begin{equation}
R_{\Xi_c}^{\mu e}\simeq 0.976.
\label{eq:RXi_me_num}
\end{equation}
The proximity of $R_{\Xi_c}^{\mu e}$ to unity reflects the suppression
of light-lepton mass effects in the Standard Model. At the same time,
the small but nonvanishing deviation from unity provides the kinematic
sensitivity required to probe lepton-nonuniversal contributions in the
charged-current sector.
\begin{figure}[t]
	\centering
	\includegraphics[width=0.48\textwidth]{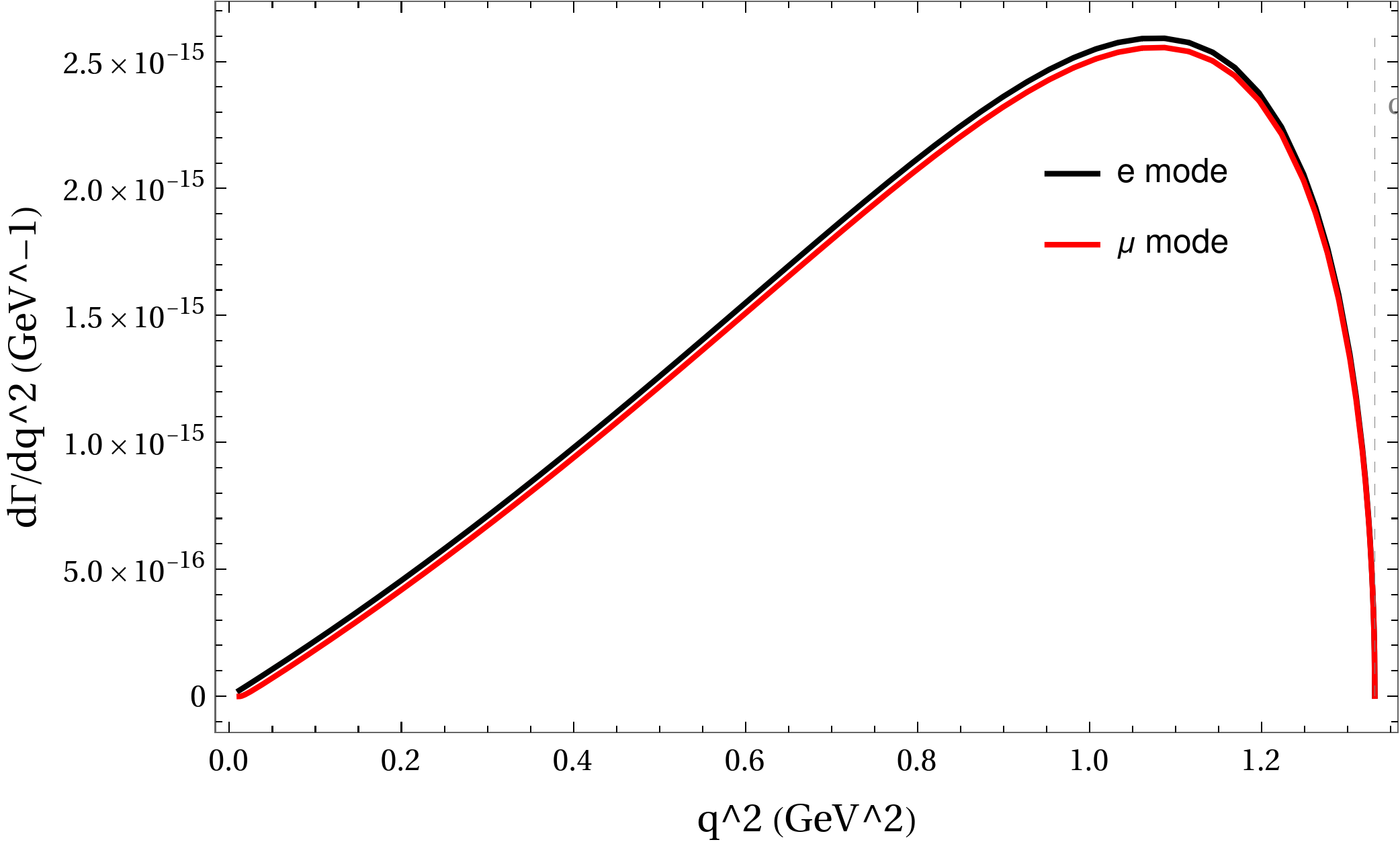}
	\caption{Differential decay rates for the semileptonic transitions
		$\Xi_{cc}\to \Xi_c e\nu$ (black) and $\Xi_{cc}\to \Xi_c \mu\nu$ (red),
		obtained using a dipole parametrization of the leading form factors.
		The small suppression of the muon mode relative to the electron mode
		arises from finite lepton-mass effects and leads to the benchmark
		prediction $R_{\Xi_c}^{\mu e}\simeq 0.976$.}
	\label{fig:dGamma_mue}
\end{figure}
\subsection{Physics interpretation}
The central feature of the observable $R_{\Xi_c}^{\mu e}$ is the
cancellation of the leading hadronic normalization between the two
decay modes, which renders the ratio insensitive to the overall
magnitude of baryonic form factors at leading order. As a result, the
observable is dominantly controlled by phase-space effects and by the
Lorentz structure of the underlying charged-current interaction.
The Standard Model prediction
$R_{\Xi_c}^{\mu e}\simeq 0.976$ reflects the suppression of
light-lepton mass effects, with the deviation from unity arising
entirely from the finite muon mass. This establishes a well-defined
reference point, against which potential deviations induced by
nonuniversal charged-current interactions can be assessed.
From a broader perspective, semileptonic doubly charmed baryon decays
provide a complementary probe of short-distance dynamics in a hadronic
environment with a distinct symmetry structure compared to mesonic
systems. The reduced sensitivity to hadronic normalization, together
with the controlled kinematic origin of the Standard Model deviation,
makes $R_{\Xi_c}^{\mu e}$ a theoretically robust observable for
probing lepton-nonuniversal effects.
In particular, the structure of the ratio allows a direct interpretation
in terms of effective operators, where deviations from the Standard
Model can be mapped onto Wilson coefficients of semileptonic
interactions. This connection is developed in the next section.
\section{Effective field theory interpretation}
\label{sec:eft}
The observable $R_{\Xi_c}^{\mu e}$ provides direct access to the
short-distance structure of the charged-current interaction. To
systematically parameterize possible deviations from the Standard Model
(SM), we employ a low-energy effective field theory (EFT) description.
At scales below the electroweak scale, semileptonic transitions are
described by the effective Lagrangian~\cite{Buchmuller:1985jz,Grzadkowski:2010es,Cirigliano:2009wk,Gonzalez-Alonso:2018omy}
\begin{equation}
\mathcal{L}_{\rm eff}
=
\mathcal{L}_{\rm SM}
+
\sum_i \frac{C_i}{\Lambda^2}\,\mathcal{O}_i,
\end{equation}
where the Wilson coefficients $C_i$ encode the effects of heavy degrees of freedom integrated out at the scale $\Lambda$; we follow standard EFT conventions for semileptonic charged-current operators~\cite{Aebischer:2017ugx}.
A central feature of the LFU ratio $R_{\Xi_c}^{\mu e}$ is the
cancellation of the leading hadronic normalization between the two
decay channels. As a consequence, the observable is predominantly
sensitive to \emph{relative} modifications of the muon and electron
amplitudes, rather than to their overall normalization. This property
allows for a transparent mapping between EFT operator structures and
observable effects.
In this work, we focus on the operators that contribute most directly
to the charged-current amplitude,
\begin{align}
\mathcal{O}_{V_L}
&=
(\bar q \gamma^\mu P_L c)(\bar \ell \gamma_\mu P_L \nu),
\\
\mathcal{O}_{S_L}
&=
(\bar q P_L c)(\bar \ell P_L \nu),
\end{align}
which capture, respectively, vector and scalar deformations of the SM
interaction.
Their impact can be directly propagated to the semileptonic amplitudes
and hence to the LFU ratio
\begin{equation}
R_{\Xi_c}^{\mu e}
=
\frac{\Gamma(\Xi_{cc} \to \Xi_c \mu \nu)}
{\Gamma(\Xi_{cc} \to \Xi_c e \nu)}.
\end{equation}
\subsection{Response to left-handed vector interactions}

We analyze the response of $R_{\Xi_c}^{\mu e}$ to deformations of the
left-handed vector operator $\mathcal O_{V_L}$ introduced above.

\begin{figure}[t]
	\centering
	\includegraphics[width=0.48\textwidth]{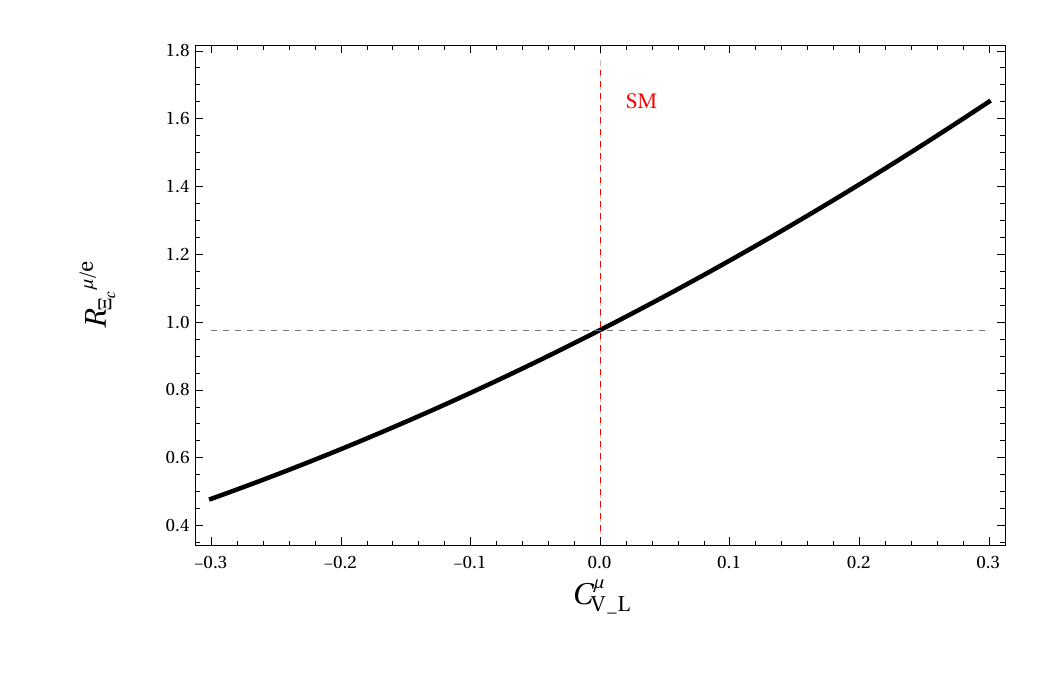}
	\caption{
		Dependence of the light-lepton universality ratio $R_{\Xi_c}^{\mu e}$
		on a lepton-nonuniversal left-handed Wilson coefficient
		$C_{V_L}^{\mu}$. The vertical dashed line indicates the Standard
		Model point, while the horizontal dashed line corresponds to the SM
		prediction. The strong variation of the ratio reflects the enhanced
		sensitivity of $\Xi_{cc}\to \Xi_c \ell \nu$ decays to
		lepton-flavor nonuniversality in vector interactions.
	}\label{fig:RXic_CVLmu}
\end{figure}

\paragraph{Universal deformation.}
For a lepton-universal modification of the charged current,
\begin{equation}
C_{V_L}^e = C_{V_L}^\mu \equiv C_{V_L},
\end{equation}
the semileptonic amplitudes are uniformly rescaled,
\begin{equation}
\mathcal A_\ell \to (1+C_{V_L})\,\mathcal A_\ell,
\qquad
\Gamma_\ell \to (1+C_{V_L})^2\,\Gamma_\ell.
\end{equation}
As a consequence, the common normalization factor cancels identically in
the LFU ratio
\begin{equation}
R_{\Xi_c}^{\mu e}(C_{V_L})
=
\frac{(1+C_{V_L})^2\,\Gamma_\mu}{(1+C_{V_L})^2\,\Gamma_e}
\simeq
R_{\Xi_c}^{\mu e}(0).
\label{eq:RXi_CVL_cancel}
\end{equation}
Numerically, we find
\begin{equation}
R_{\Xi_c}^{\mu e}(C_{V_L}=0)=0.9759,
\qquad
R_{\Xi_c}^{\mu e}(C_{V_L}=\pm 0.1)\simeq 0.9759,
\end{equation}
confirming that universal left-vector deformations cancel in the
observable to excellent approximation. The ratio is therefore
insensitive to an overall normalization shift of the charged-current
amplitude.
\paragraph{Lepton-nonuniversal deformation.}
A qualitatively different behavior arises if lepton universality is
violated. We consider
\begin{equation}
C_{V_L}^{\mu}\neq 0,
\qquad
C_{V_L}^{e}=0,
\end{equation}
so that only the muon channel is modified,
\begin{equation}
\mathcal A_{\mu}\to (1+C_{V_L}^{\mu})\,\mathcal A_{\mu},
\qquad
\mathcal A_{e}\to \mathcal A_{e}.
\end{equation}
The LFU ratio then follows the scaling relation
\begin{equation}
R_{\Xi_c}^{\mu e}(C_{V_L}^{\mu})
=
(1+C_{V_L}^{\mu})^2\,R_{\Xi_c}^{\mu e}(0).
\label{eq:RXi_nonuniv}
\end{equation}
This quadratic scaling is exact at leading order and reflects the
absence of additional operator interference structures in the
light-lepton limit.
Using the benchmark form-factor parametrization introduced above, we obtain
\begin{equation}
R_{\Xi_c}^{\mu e}(C_{V_L}^{\mu}=0)=0.9759,
\end{equation}
\begin{equation}
R_{\Xi_c}^{\mu e}(C_{V_L}^{\mu}=0.1)\simeq 1.1808,
\qquad
R_{\Xi_c}^{\mu e}(C_{V_L}^{\mu}=-0.1)\simeq 0.7904.
\end{equation}
The observable is therefore highly sensitive to lepton-nonuniversal
vector interactions, while remaining essentially blind to universal
vector rescalings.
The dependence shown in Fig.~\ref{fig:RXic_CVLmu} makes this structure
transparent. Values $C_{V_L}^{\mu}\sim \mathcal O(10^{-1})$ induce
order-one shifts in $R_{\Xi_c}^{\mu e}$, reflecting the direct coupling
of the operator to the dominant Standard Model amplitude. Combined with
the cancellation of leading hadronic normalization, this establishes
$R_{\Xi_c}^{\mu e}$ as a theoretically robust probe of
lepton-flavor nonuniversality in the charged-current sector.
\subsection{Operator sensitivity in semileptonic $\Xi_{cc}$ decays}
\label{sec:operator_sensitivity}
To quantify the sensitivity of the observable $R_{\Xi_c}^{\mu e}$ to
short-distance physics, we compare the impact of lepton-nonuniversal
vector and scalar deformations within the EFT framework introduced
above~\cite{Buchmuller:1985jz,Grzadkowski:2010es,Cirigliano:2009wk}.
We consider benchmark scenarios in which only the muon channel is
modified,
\begin{equation}
C_{V_L}^\mu \neq 0, \qquad C_S^\mu \neq 0,
\qquad C^e = 0.
\end{equation}
\paragraph{Scalar contributions.}
Scalar interactions enter the decay rate through helicity-suppressed
terms proportional to the lepton mass,
\begin{equation}
\delta \Gamma_S \propto \frac{m_\mu^2}{q^2}\,|C_S^\mu|^2,
\end{equation}
which suppresses their effect in the light-lepton regime.
Using the same form-factor input as above, we obtain
\begin{equation}
R_{\Xi_c}^{\mu e}(C_S^\mu=0)=0.9759,
\end{equation}
\begin{equation}
R_{\Xi_c}^{\mu e}(C_S^\mu=0.1)\simeq 0.9781,
\qquad
R_{\Xi_c}^{\mu e}(C_S^\mu=-0.1)\simeq 0.9738.
\end{equation}
Even for benchmark Values $C_{V_L}^{\mu}\sim \mathcal{O}(10^{-1})$ are used here as
illustrative benchmarks to demonstrate the parametric response of the
observable, while the projected experimental sensitivity corresponds to
$|C_{V_L}^{\mu}|\sim \mathcal{O}(10^{-2})$. the deviation remains at the
percent level, demonstrating that $R_{\Xi_c}^{\mu e}$ is parametrically insensitive to scalar interactions.

\begin{figure}[t]
	\centering
	\includegraphics[width=0.48\textwidth]{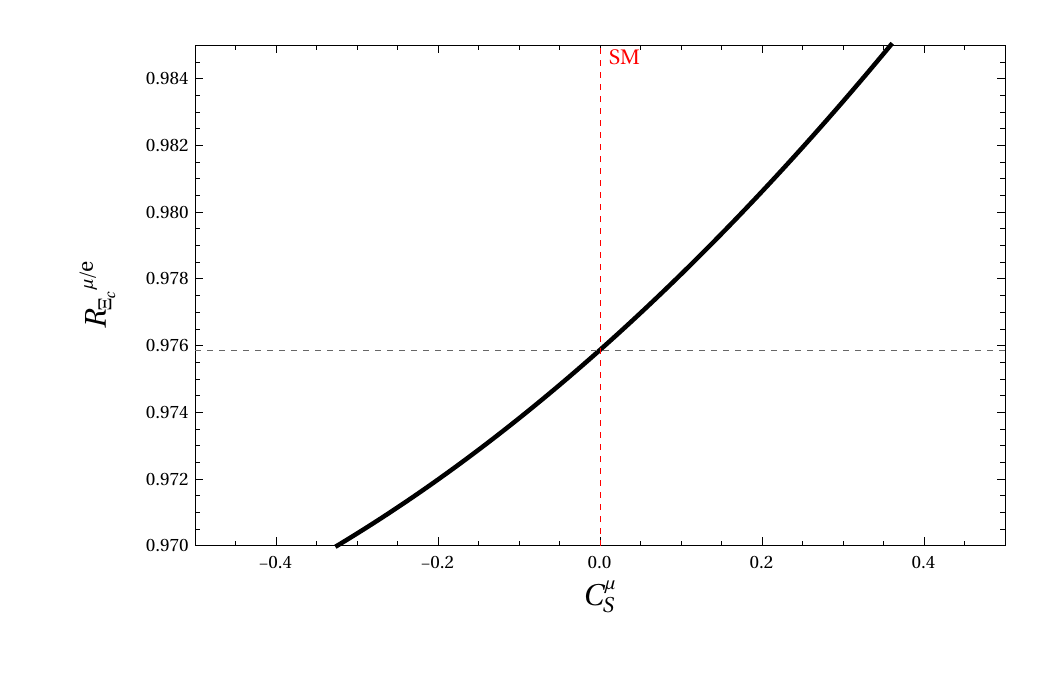}
	\caption{
		Dependence of $R_{\Xi_c}^{\mu e}$ on the scalar Wilson coefficient
		$C_S^\mu$ affecting only the muon channel.
		The restricted vertical scale highlights the parametric suppression
		of scalar contributions arising from the $m_\mu^2/q^2$ factor.
	}
	\label{fig:RXicCSmu}
\end{figure}

\begin{figure}[htbp]
	\centering
	\includegraphics[width=0.48\textwidth]{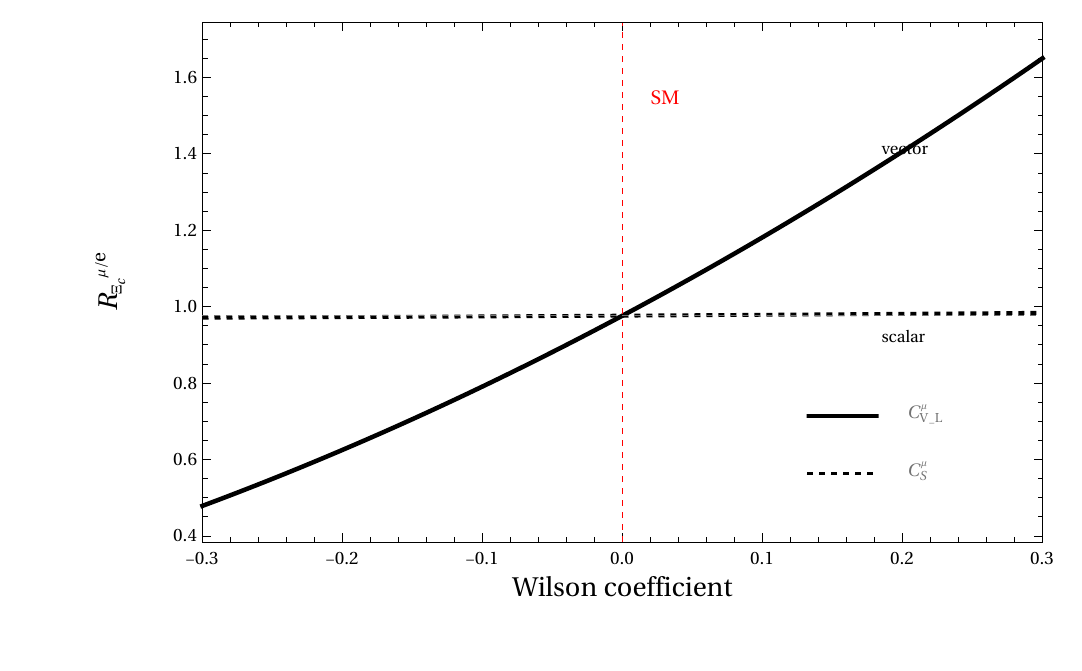}
	\caption{
		Comparison of the dependence of $R_{\Xi_c}^{\mu e}$ on a
		vector coefficient $C_{V_L}^\mu$ (solid) and a scalar coefficient
		$C_S^\mu$ (dashed).
		Vector interactions induce order-one deviations, while scalar
		contributions remain confined to percent-level shifts due to
		helicity suppression.
	}
	\label{fig:RXic_compare}
\end{figure}

This behavior is illustrated in Fig.~\ref{fig:RXicCSmu}, where the
dependence on $C_S^\mu$ is confined to a narrow band around the Standard
Model prediction.
\paragraph{Vector contributions.}
A left-handed vector deformation rescales the Standard Model amplitude,
\begin{equation}
\mathcal A_\mu \to (1 + C_{V_L}^\mu)\,\mathcal A_\mu,
\end{equation}
leading to
\begin{equation}
R_{\Xi_c}^{\mu e}(C_{V_L}^\mu)
=
(1 + C_{V_L}^\mu)^2\,R_{\Xi_c}^{\mu e}(\mathrm{SM}).
\end{equation}
This unsuppressed interference produces an $\mathcal{O}(1)$ response
even for moderate values of $C_{V_L}^\mu$,
\begin{equation}
R_{\Xi_c}^{\mu e}(C_{V_L}^\mu = 0.1) \simeq 1.18,
\qquad
R_{\Xi_c}^{\mu e}(\mathrm{SM}) \simeq 0.976.
\end{equation}
\paragraph{Hierarchy and discrimination.}
The comparison reveals a pronounced hierarchy: vector interactions
generate order-one deviations through direct interference with the SM
current, whereas scalar contributions are suppressed by
$m_\mu^2/q^2$ and remain subleading.
This behavior is summarized in Fig.~\ref{fig:RXic_compare}.
The observable $R_{\Xi_c}^{\mu e}$ therefore acts as a direct probe of
lepton-flavor nonuniversal vector interactions, while being effectively
insensitive to scalar new physics at the level of current experimental
precision. The cancellation of leading hadronic normalization ensures
that this discrimination is driven by short-distance dynamics rather
than by nonperturbative inputs~\cite{Tanaka:2010se,Sakaki:2013bfa,Ligeti:2016npd}.
\subsection{Phenomenological hierarchy and interpretation}
The analysis above establishes a well-defined hierarchy in the
sensitivity of $R_{\Xi_c}^{\mu e}$ to different operator structures.
Universal left-handed vector interactions cancel in the ratio due to
their common rescaling of the semileptonic amplitudes, rendering the
observable insensitive to overall normalization shifts of the charged
current. In contrast, lepton-nonuniversal vector interactions induce a
quadratic enhancement through unsuppressed interference with the
Standard Model amplitude, leading to order-one deviations for
$\mathcal O(10^{-1})$ Wilson coefficients. Scalar contributions, by
comparison, are parametrically suppressed by the light-lepton mass and
therefore generate only percent-level effects in the same parameter
range.
This hierarchy follows directly from the Lorentz structure of the
effective operators and the kinematics of light-lepton final states.
Importantly, it is largely independent of the detailed hadronic
description, as the leading baryonic normalization cancels in the
observable. The ratio $R_{\Xi_c}^{\mu e}$ thus isolates short-distance
physics in a controlled and theoretically robust manner, with a
parametric sensitivity that cleanly distinguishes between operator
classes.
\subsection{Complementarity with nonleptonic observables}
A central outcome of this framework is the complementarity between the
semileptonic ratio $R_{\Xi_c}^{\mu e}$ and the nonleptonic null
observable $\Delta_{\pi K}$. The latter is constructed to vanish in the
heavy-diquark factorization limit and therefore provides a direct probe
of nonfactorizable QCD dynamics. In contrast, $R_{\Xi_c}^{\mu e}$ is
protected from leading hadronic normalization effects and is primarily
sensitive to short-distance operator structure.
The two observables therefore probe orthogonal aspects of the dynamics.
A deviation in $\Delta_{\pi K}$ signals the presence of subleading QCD
effects, while a deviation in $R_{\Xi_c}^{\mu e}$ points to
lepton-flavor nonuniversality or nonstandard operator contributions.
The simultaneous analysis of both quantities thus provides a powerful
consistency test: a purely hadronic effect cannot reproduce the
operator-dependent scaling observed in $R_{\Xi_c}^{\mu e}$, while a
short-distance deformation alone does not generically generate a
nonvanishing $\Delta_{\pi K}$.
This interplay establishes doubly charmed baryons as a uniquely
sensitive laboratory in which QCD dynamics and short-distance physics
can be disentangled in a controlled and complementary manner.
\section{Prospective constraints on new physics}
\label{sec:constraints}
\begin{figure}[t]
	\centering
	\includegraphics[width=0.48\textwidth]{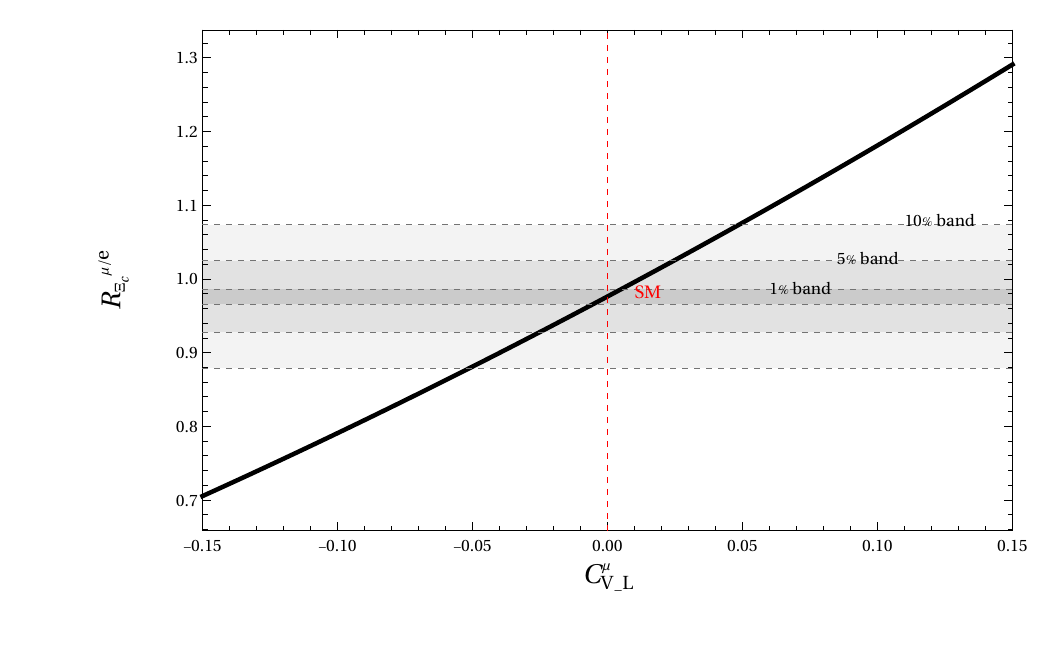}
	\caption{Projected sensitivity of the light-lepton universality ratio $R_{\Xi_c}^{\mu e}$ to a lepton-nonuniversal vector Wilson coefficient $C_{V_L}^{\mu}$. The vertical dashed line marks the Standard Model point, while the horizontal shaded bands indicate hypothetical future measurements centered on the SM prediction with relative precisions of $1\%$, $5\%$, and $10\%$. The intersections of these bands with the theory curve determine the corresponding allowed intervals for $C_{V_L}^{\mu}$.}
	\label{fig:projCVLmu}
\end{figure}

The observable $R_{\Xi_c}^{\mu e}$ provides a theoretically clean
probe of lepton-flavor nonuniversality in charged-current interactions.
Due to the cancellation of leading hadronic uncertainties and its
simple dependence on short-distance Wilson coefficients, it enables
a direct and quantitatively robust interpretation in terms of
constraints on effective operators.
Figure~\ref{fig:projCVLmu} shows the predicted dependence of $R_{\Xi_c}^{\mu e}$ on $C_{V_L}^{\mu}$ together with benchmark experimental sensitivity bands corresponding to relative precisions of $1\%$, $5\%$, and $10\%$ around the Standard Model prediction. Since the ratio obeys the exact scaling relation
\begin{equation}
R_{\Xi_c}^{\mu e}(C_{V_L}^{\mu})
=
(1+C_{V_L}^{\mu})^2\,R_{\Xi_c}^{\mu e}(0),
\end{equation}
the projected bounds can be obtained analytically.

The parameter $\epsilon_{\rm exp}$ denotes the relative experimental
precision on the measurement of $R_{\Xi_c}^{\mu e}$. It is defined as
\begin{equation}
R_{\Xi_c}^{\mu e,\rm exp}
=
R_{\Xi_c}^{\mu e}(0)\,(1 \pm \epsilon_{\rm exp}),
\end{equation}
and does not arise from theoretical uncertainties, but parametrizes the
expected experimental sensitivity.

For an experimental determination consistent with the Standard Model
prediction with relative precision $\epsilon_{\rm exp}$, one finds
\begin{equation}
\sqrt{1-\epsilon_{\rm exp}}-1
\le
C_{V_L}^{\mu}
\le
\sqrt{1+\epsilon_{\rm exp}}-1.
\end{equation}

This implies the benchmark sensitivity ranges
\begin{align}
1\%:&\qquad -5.0\times10^{-3}\lesssim C_{V_L}^{\mu}\lesssim 5.0\times10^{-3},
\\
5\%:&\qquad -2.53\times10^{-2}\lesssim C_{V_L}^{\mu}\lesssim 2.47\times10^{-2},
\\
10\%:&\qquad -5.13\times10^{-2}\lesssim C_{V_L}^{\mu}\lesssim 4.88\times10^{-2}.
\end{align}
These results show that even a moderate measurement of $R_{\Xi_c}^{\mu e}$ at the few-percent level would already probe light-lepton flavor nonuniversality at the level of a few times $10^{-2}$ in the vector coefficient. This analytic mapping demonstrates that $R_{\Xi_c}^{\mu e}$ directly translates percent-level experimental precision into sensitivity to new physics at the level $|C_{V_L}^{\mu}|\sim \mathcal{O}(10^{-2})$.
Remarkably, this places the baryonic observable in the same parametric regime currently probed by semileptonic $B$-meson measurements~\cite{Fajfer:2012vx,Bernlochner:2017jka}, establishing it as a quantitatively competitive probe of charged-current new physics.
\begin{figure}[t]
	\centering
	\includegraphics[width=0.48\textwidth]{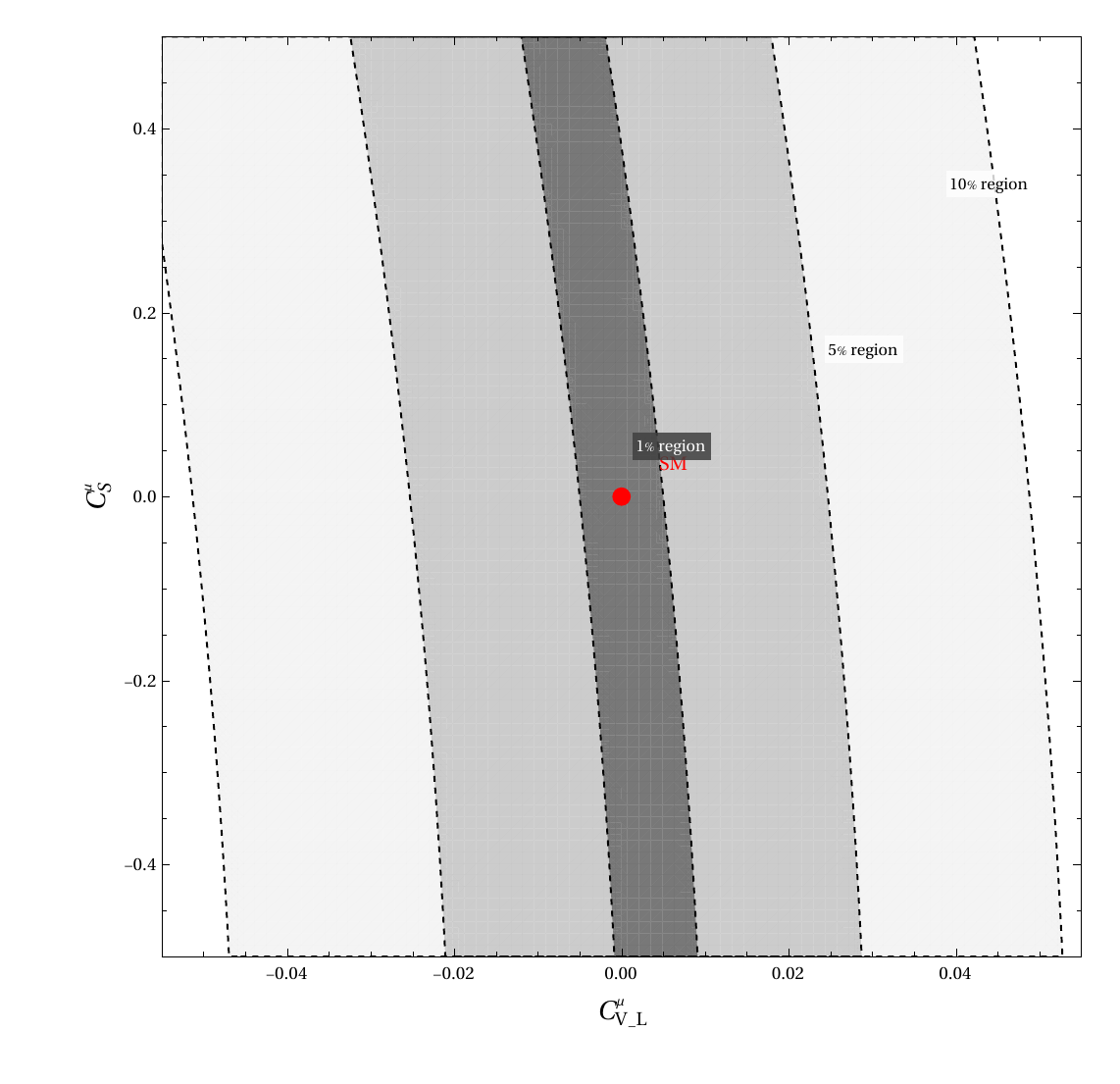}
	\caption{Projected allowed regions in the $(C_{V_L}^{\mu},\,C_S^{\mu})$ plane
		derived from hypothetical measurements of the ratio
		$R_{\Xi_c}^{\mu e}$ with relative precisions of $1\%$, $5\%$, and $10\%$,
		shown as progressively wider bands around the Standard Model point.
		The allowed regions exhibit a nearly vertical structure, indicating a strong sensitivity of the observable to lepton-nonuniversal vector interactions, while remaining largely insensitive to scalar contributions.This geometric feature directly reflects the parametric hierarchy between vector and scalar operators in light-lepton semileptonic decays, where scalar effects are suppressed by the lepton mass.}\label{fig:allowed2D}
\end{figure}
The corresponding sensitivity to scalar operators is parametrically weaker. Because scalar contributions are helicity-suppressed by the muon mass, the same experimental precision translates into significantly weaker bounds on $C_S^\mu$. In this sense, $R_{\Xi_c}^{\mu e}$ should primarily be viewed as a precision probe of lepton-nonuniversal vector interactions, with scalar sensitivity playing a secondary and complementary role.
The two-dimensional constraint shown in Fig.~\ref{fig:allowed2D}
exhibits a pronounced directional structure in the space of Wilson
coefficients. The allowed regions are strongly elongated along the
$C_S^{\mu}$ direction, while remaining tightly constrained in
$C_{V_L}^{\mu}$. This reflects the underlying helicity structure of
the effective operators: vector interactions interfere linearly with the Standard Model amplitude and are not helicity suppressed, leading to an unsuppressed parametric
contribution to the decay rate. whereas scalar contributions are suppressed by $m_\mu^2/q^2$ and enter only at subleading order. This directional sensitivity implies that $R_{\Xi_c}^{\mu e}$ can effectively disentangle vector and scalar operator contributions,
providing information that is orthogonal to that obtained from mesonic observables where such degeneracies are more pronounced.
This highlights the role of baryonic observables as a powerful tool
for resolving operator degeneracies in global analyses.
As a consequence, $R_{\Xi_c}^{\mu e}$ acts as a highly selective probe of lepton-nonuniversal vector interactions, This directional sensitivity shows that $R_{\Xi_c}^{\mu e}$ primarily constrains lepton-nonuniversal vector interactions, while scalar directions remain comparatively weakly constrained because of helicity
suppression.
\subsection{Consistency with existing light-lepton meson constraints}
\label{sec:meson_consistency}
\begin{figure}[htbp]
	\centering
	\includegraphics[width=0.48\textwidth]{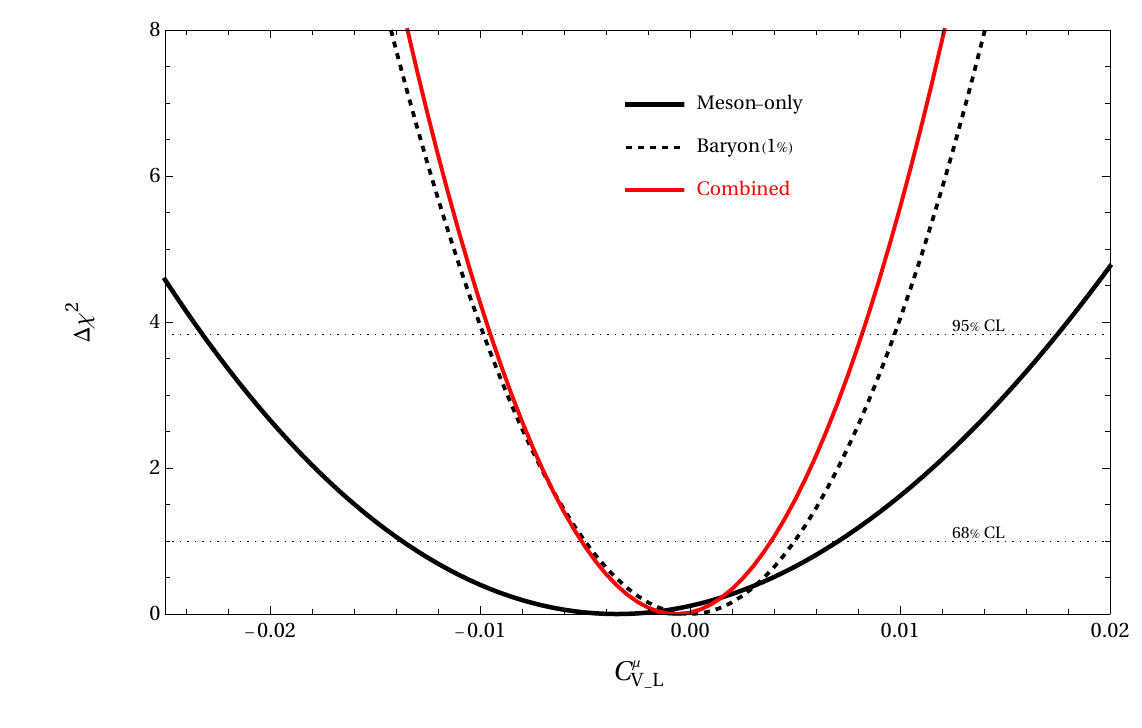}
	\caption{
		$\Delta\chi^2$ profiles for the Wilson coefficient
		$C_{V_L}^{\mu}$. The black solid curve corresponds to the
		constraint derived from the inclusive meson ratio
		$R(X_{e/\mu})$, while the dashed curve shows the projected
		sensitivity of the baryonic observable
		$R_{\Xi_c}^{\mu e}$ assuming a $1\%$ measurement centered on
		the Standard Model prediction. The red curve represents the
		combined fit. Horizontal lines indicate the $68\%$ and $95\%$
		confidence levels. The combination significantly reduces the
		allowed parameter space and pulls the preferred value toward
		the Standard Model point. }\label{fig:DeltaChi2_CVLmu}
\end{figure}

A central question is whether the baryonic observable
$R_{\Xi_c}^{\mu e}$ probes a region of parameter space that is already
constrained by existing measurements of light-lepton universality.
Addressing this issue is essential to assess whether semileptonic doubly charmed baryon
decays provide independent and complementary constraints on charged-current new physics beyond those obtained from mesonic observables. A particularly clean input is provided by the inclusive ratio
\begin{equation}
R(X_{e/\mu}) =
\frac{\mathcal B(B \to X e \nu_e)}
{\mathcal B(B \to X \mu \nu_\mu)},
\end{equation}
which is largely insensitive to hadronic form-factor uncertainties and
therefore constitutes a robust test of light-lepton universality. The
Belle II Collaboration reports
\begin{equation}
R(X_{e/\mu}) = 1.007 \pm 0.021,
\end{equation}
where the quoted uncertainty includes both statistical and systematic
contributions~\cite{BelleII:2023RXemu}.
Within the effective-field-theory framework considered here, and
assuming that new physics affects only the muon channel, this observable
scales as
\begin{equation}
R(X_{e/\mu})(C_{V_L}^{\mu})
\simeq
\frac{1}{(1 + C_{V_L}^{\mu})^2},
\end{equation}
while the baryonic ratio follows
\begin{equation}
R_{\Xi_c}^{\mu e}(C_{V_L}^{\mu})
=
(1 + C_{V_L}^{\mu})^2\,R_{\Xi_c}^{\mu e}(0).
\end{equation}
This complementary dependence highlights that the two observables probe
the same Wilson coefficient with opposite parametric behavior.
This opposite scaling is phenomenologically useful because it makes the
baryonic observable a consistency test rather than a simple repetition
of the mesonic constraint. A common value of $C_{V_L}^{\mu}$ must
simultaneously account for both the inverse response of
$R(X_{e/\mu})$ and the direct response of $R_{\Xi_c}^{\mu e}$.
To quantify their interplay, we construct a $\chi^2$ function for the
meson constraint,
\begin{equation}
\chi^2_{\rm meson}(C_{V_L}^{\mu}) =
\frac{\left[R(X_{e/\mu})(C_{V_L}^{\mu}) - R_{\rm exp}\right]^2}
{\sigma^2},
\end{equation}
and combine it with the projected baryonic contribution
\begin{equation}
\chi^2_{\rm baryon}(C_{V_L}^{\mu}) =
\frac{\left[R_{\Xi_c}^{\mu e}(C_{V_L}^{\mu}) -
	R_{\Xi_c}^{\mu e}(0)\right]^2}
{(\epsilon_{\rm exp}\,R_{\Xi_c}^{\mu e}(0))^2},
\end{equation}
where $\epsilon_{\rm exp}$ denotes the assumed relative experimental precision of the projected baryonic measurement. This framework enables a direct statistical comparison between existing meson data and projected baryonic measurements within a common EFT
description, thereby providing a nontrivial consistency test of short-distance charged-current interactions. The Standard Model reference value entering the baryonic
contribution to the $\chi^2$ function, $R_{\Xi_c}^{\mu e}(0) \simeq 0.976$, is taken from the
phase-space integration described in Sec.~\ref{sec:semileptonic}. Figure~\ref{fig:DeltaChi2_CVLmu} shows the resulting $\Delta\chi^2$ profiles for the meson constraint, the projected baryonic sensitivity at the $1\%$ level, and their combination.
The meson-only profile exhibits a mild preference for a small negative value of $C_{V_L}^{\mu}$, reflecting the current experimental uncertainty. In contrast, the projected baryonic measurement imposes a symmetric constraint around $C_{V_L}^{\mu}=0$, as it is assumed to be centered on the Standard Model prediction. The combination of the two observables leads to a substantial reduction of the allowed parameter space and shifts the best-fit point toward the Standard Model. Importantly, this improvement arises from the inclusion of an independent observable probing the same Wilson coefficient in a distinct hadronic environment, rather than from increased statistical precision alone. A key outcome of this analysis is that percent-level measurements of $R_{\Xi_c}^{\mu e}$ probe the same Wilson coefficient with a parametric sensitivity that is orthogonal to mesonic observables. As a result, As a result, the baryonic input does not only improve precision, but qualitatively alters the structure of the constraint by lifting degeneracies inherent to meson-only fits.
This demonstrates that semileptonic doubly charmed baryon decays provide a genuinely independent and quantitatively competitive probe of lepton-nonuniversal charged-current interactions. In particular, a percent-level measurement of $R_{\Xi_c}^{\mu e}$ would enable a meaningful consistency test of existing light-lepton universality constraints.
A deviation between the baryonic measurement and the meson constraint would signal either a breakdown of the simple vector-like EFT scenario or the presence of additional operator structures, while agreement would provide a stringent cross-check of the Standard Model description in a hadronic system governed by distinct nonperturbative QCD dynamics.
\subsection{Evolution of constraints with baryonic precision}
\label{sec:constraint_evolution}
\begin{figure}[t]
	\centering
	\includegraphics[width=0.48\textwidth]{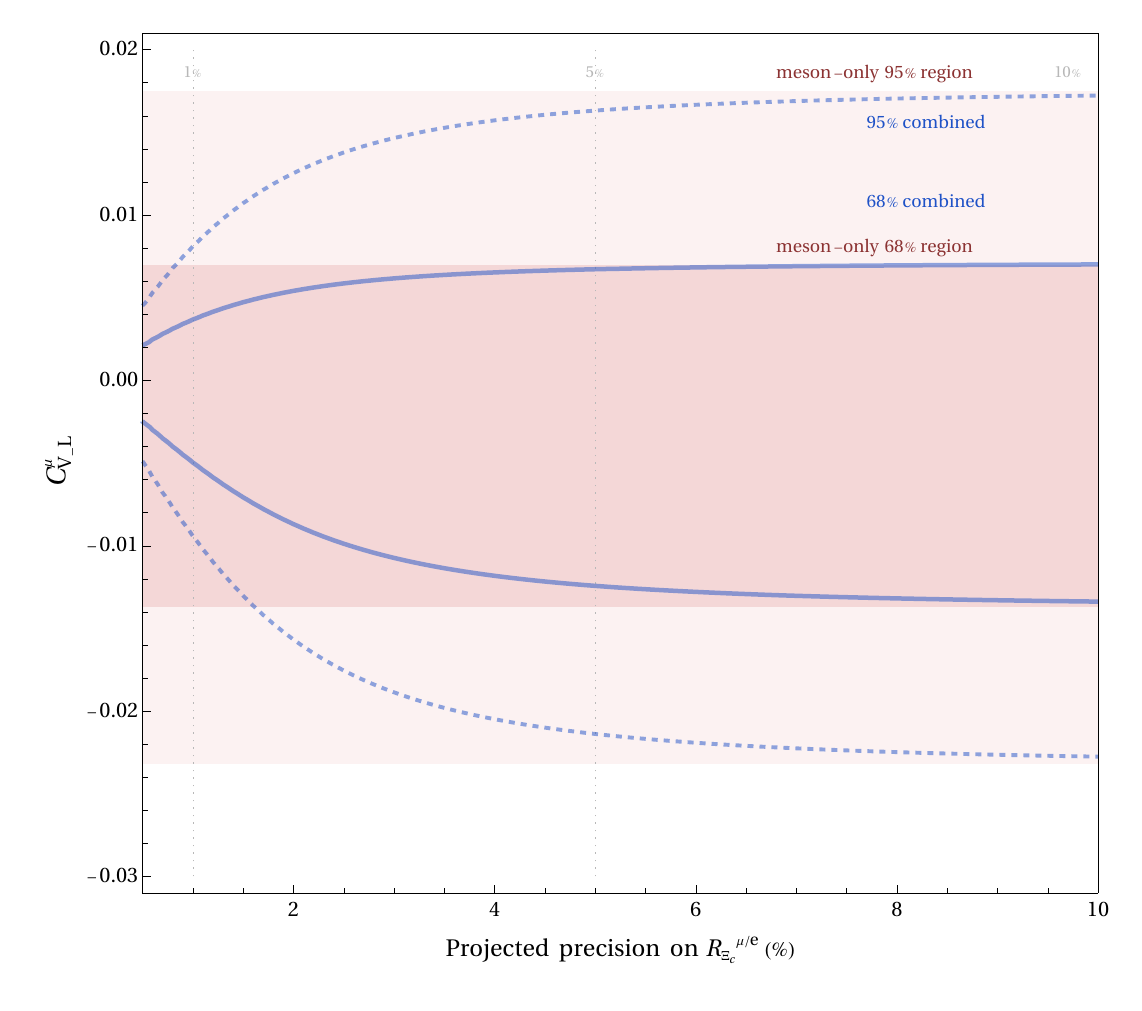}
	\caption{
		Evolution of the allowed region for the Wilson coefficient
		$C_{V_L}^{\mu}$ as a function of the projected experimental
		precision on $R_{\Xi_c}^{\mu e}$. The horizontal shaded bands
		indicate the $68\%$ and $95\%$ confidence intervals derived from
		the Belle II measurement of the inclusive ratio
		$R(X_{e/\mu})$, while the curves show the corresponding intervals
		obtained from the combined meson+baryon fit. The complementary
		scaling of the two observables leads to a progressive reduction
		of the allowed parameter space as the baryonic precision improves.
		A percent-level measurement of $R_{\Xi_c}^{\mu e}$ yields a
		significant tightening of the constraint, demonstrating the
		competitive sensitivity of semileptonic $\Xi_{cc}$ decays to
		lepton-nonuniversal vector interactions.
	}
	\label{fig:ConstraintEvolution_CVLmu}
\end{figure}

Having established the complementarity between mesonic constraints and baryonic observables, we now determine the precision at which semileptonic $\Xi_{cc}$ decays begin to provide quantitatively competitive constraints on the Wilson coefficient $C_{V_L}^{\mu}$.
To this end, we extract the allowed intervals for the Wilson coefficient $C_{V_L}^{\mu}$ at the $68\%$ and $95\%$ confidence levels as a function of the assumed relative precision $\epsilon_{\rm exp}$ on $R_{\Xi_c}^{\mu e}$, combining the Belle II constraint on $R(X_{e/\mu})$ with the projected baryonic contribution to the $\chi^2$ function. The result is shown in Fig.~\ref{fig:ConstraintEvolution_CVLmu}. The horizontal shaded bands represent the meson-only constraints, while the curves show the corresponding allowed regions obtained from the combined fit as the baryonic precision is varied. The behavior is governed by the complementary parametric dependence of the two observables,
\[
R(X_{e/\mu}) \propto (1 + C_{V_L}^{\mu})^{-2}, \qquad
R_{\Xi_c}^{\mu e} \propto (1 + C_{V_L}^{\mu})^{2},
\]
which ensures that the baryonic input constrains the same Wilson
coefficient with opposite scaling. As a result, the inclusion of
baryonic information efficiently lifts the degeneracy present in the
meson-only constraint.
Quantitatively, the meson-only measurement allows for
$\mathcal{O}(10^{-2})$ deviations in $C_{V_L}^{\mu}$, reflecting the
current experimental uncertainty on $R(X_{e/\mu})$. The addition of a
baryonic measurement leads to a rapid contraction of the allowed
parameter space, becoming significant already at the few-percent level
and pronounced once $\epsilon_{\rm exp} \sim 1\%$.
In particular, a percent-level determination of
$R_{\Xi_c}^{\mu e}$ reduces the $68\%$ confidence interval on
$C_{V_L}^{\mu}$ by more than a factor of two, with a comparable
contraction of the $95\%$ region. This establishes that semileptonic
$\Xi_{cc}$ decays impose constraints on lepton-nonuniversal
charged-current interactions at a similar level of sensitivity to
light-lepton meson measurements, while accessing complementary hadronic
dynamics.
More importantly, this improvement does not arise from increased
statistics alone, but from the inclusion of an independent observable probing
the same short-distance interaction in a qualitatively different hadronic system. This establishes a genuine consistency test of the effective-field-theory description.
The figure therefore identifies the precision threshold at which
baryonic observables transition from a complementary probe to a
quantitatively competitive constraint on $C_{V_L}^{\mu}$.
\subsection{Universal left-vector benchmark across light- and tau-lepton observables}
\label{sec:universal_CVL}
\begin{figure}[htbp]
	\centering
	\includegraphics[width=0.48\textwidth]{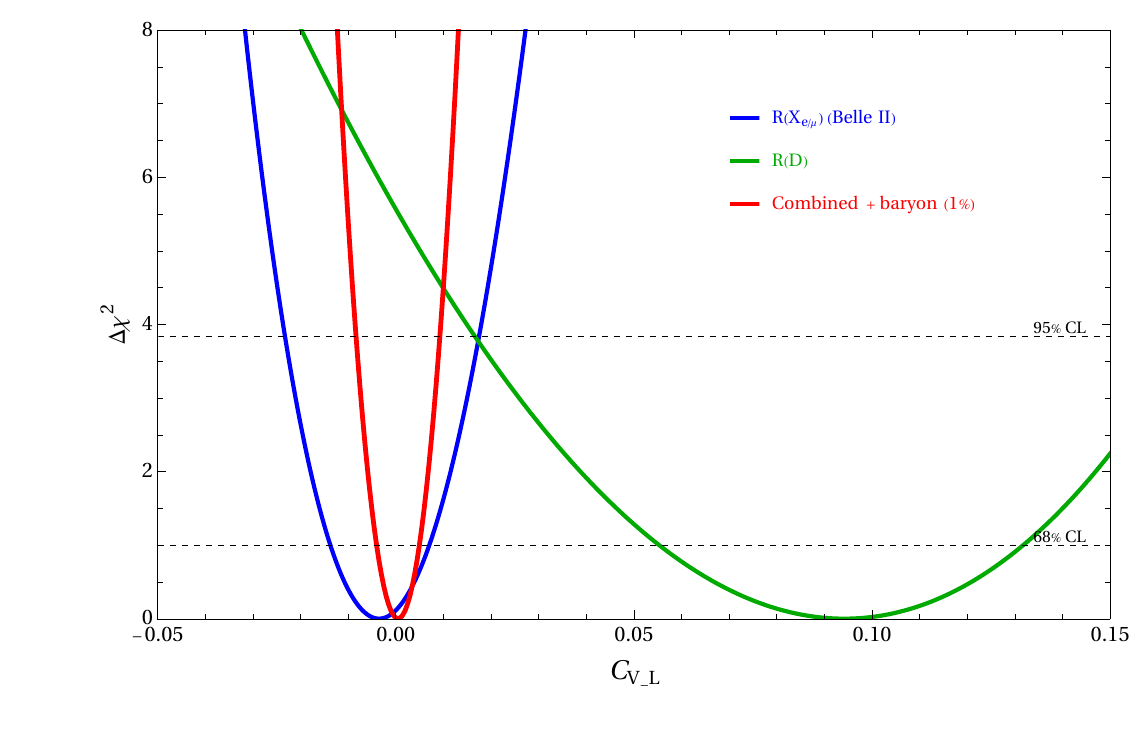}
	\caption{
		$\Delta\chi^2$ profiles for a universal left-vector Wilson coefficient
		$C_{V_L}$, assuming $C_{V_L}^{\mu} = C_{V_L}^{\tau}$.
		The blue curve shows the constraint from the Belle II measurement of
		the inclusive ratio $R(X_{e/\mu})$~\cite{BelleII:2023RXemu}, the green
		curve corresponds to the $R(D)$ benchmark~\cite{HFLAV:2023}, and the red
		curve shows the combined fit including a projected $1\%$ measurement of
		the baryonic observable $R_{\Xi_c}^{\mu e}$ centered on the Standard
		Model prediction.
		Horizontal dashed lines indicate the $68\%$ and $95\%$ confidence levels.
		The light-lepton and baryonic observables favor values of $C_{V_L}$
		close to zero, while $R(D)$ prefers a positive shift. The combined fit
		is therefore driven toward the Standard Model region, illustrating the
		tension that arises when attempting to describe all observables within
		a single universal left-vector deformation.
	}
	\label{fig:UniversalCVL}
\end{figure}

While the previous analysis focused on light-lepton nonuniversality,
we also consider a complementary benchmark in which a single universal
left-handed deformation modifies both light- and tau-lepton charged-current
observables.
\begin{equation}
C_{V_L}^{\mu}=C_{V_L}^{\tau}\equiv C_{V_L},
\end{equation}
This universality assumption is adopted only as a benchmark scenario,
designed to test whether a single EFT deformation can simultaneously
describe light-lepton, tau-lepton, and baryonic observables.
In this setup, the relevant observables scale as
\begin{equation}
\begin{aligned}
R(X_{e/\mu}) &\propto (1 + C_{V_L})^{-2}, \\
R(D)         &\propto (1 + C_{V_L})^{2}, \\
R_{\Xi_c}^{\mu e} &\propto (1 + C_{V_L})^{2}.
\end{aligned}
\end{equation}
where $R(X_{e/\mu})$ is measured by Belle II~\cite{BelleII:2023RXemu},
while $R(D)$ is taken as a representative semitauonic benchmark from
the HFLAV averages and global $b\to c\tau\nu$ analyses
~\cite{HFLAV:2023,Alguero:2022,Murgui:2019czp,Bernlochner:2017jka}. The observable
$R_{\Xi_c}^{\mu e}$ is the baryonic ratio introduced in this work.
To quantify the interplay between these constraints, we construct a
combined $\chi^2$ function including the Belle II measurement,
the $R(D)$ benchmark, and the projected baryonic sensitivity at the
percent level. The resulting $\Delta\chi^2$ profiles are shown in
Fig.~\ref{fig:UniversalCVL}.
The light-lepton observable $R(X_{e/\mu})$ and the projected baryonic
measurement both favor values of $C_{V_L}$ close to the Standard Model,
while the $R(D)$ benchmark prefers a positive shift of order
$C_{V_L} \sim \mathcal{O}(10^{-1})$. As a result, the combined fit is
pulled toward the Standard Model region, leading to a nontrivial tension
between the different classes of observables within this minimal
universal scenario.
The tension observed in the universal $C_{V_L}$ benchmark provides a
concrete example in which baryonic observables can reveal inconsistencies
that are less visible in meson-only projections.
This comparison highlights an important phenomenological implication.
The combined fit indicates that a single universal left-vector
deformation is disfavored as a simultaneous description of the Belle II
light-lepton constraint, the projected baryonic measurement, and the
$R(D)$ benchmark. In this sense, the baryonic observable strengthens
the case for either flavor nonuniversality or a more general
effective-field-theory structure involving additional operator
contributions beyond a simple rescaling of the Standard Model current.
\vspace{-20pt}
\section{Phenomenological implications} 
The results obtained in this work demonstrate that doubly charmed baryons provide a complementary framework for testing QCD dynamics and charged-current
interactions with distinct hadronic uncertainties. A central outcome is the identification of observables with complementary and orthogonal sensitivity to QCD dynamics and 
short-distance interactions. The nonleptonic null combination $\Delta_{\pi K}$ isolates nonfactorizable QCD effects by construction, providing a direct probe of deviations from the heavy-diquark factorization limit. In contrast, the semileptonic ratio $R_{\Xi_c}^{\mu e}$ is protected against leading hadronic uncertainties and directly probes the Lorentz structure of the charged current. The combined analysis reveals a hierarchical pattern of sensitivity. Universal vector deformations cancel identically in $R_{\Xi_c}^{\mu e}$, while lepton-nonuniversal vector interactions induce unsuppressed quadratic deviations. Scalar contributions, by contrast, remain helicity suppressed and generate only percent-level effects. This hierarchy allows a clean separation between operator classes that is largely independent of hadronic input. Importantly, the interplay between baryonic and mesonic observables provides a nontrivial consistency test of the effective-field-theory 
description. The complementary scaling of $R(X_{e/\mu})$ and $R_{\Xi_c}^{\mu e}$ lifts degeneracies present in meson-only analyses and leads to a qualitatively sharper constraint 
on the Wilson coefficient $C_{V_L}^{\mu}$. Furthermore, the universal $C_{V_L}$ benchmark highlights a tension between light-lepton, semitauonic, and baryonic observables, indicating 
that a single universal deformation cannot simultaneously describe all data. This provides evidence that viable new-physics scenarios must involve either flavor nonuniversality or additional operator structures. From an experimental perspective, the required observables are accessible at LHCb, where continued improvements in charm-baryon 
reconstruction make percent-level measurements plausible. Such measurements would directly probe Wilson coefficients at the level $|C_{V_L}^{\mu}| \sim \mathcal{O}(10^{-2})$, placing 
$\Xi_{cc}$ decays on equal footing with current mesonic probes. Taken together, these results show that $\Xi_{cc}$ decays can supply independent constraints on both long-distance QCD dynamics and short-distance charged-current interactions.
\subsection{Ultraviolet mapping and phenomenological implications}
\label{sec:UV_Wprime}
To connect the effective-field-theory analysis with explicit ultraviolet
physics, we map the constraint on $C_{V_L}^{\mu}$ onto a benchmark
charged vector boson $W'$ with left-handed couplings~\cite{Langacker:2008yv}.
Similar charged-current operators can also arise in leptoquark
realizations~\cite{Dorsner:2016wpm}, Such UV completions have been widely discussed in connection with semileptonic charged-current anomalies~\cite{Freytsis:2015qca}. although the precise mapping to collider constraints depends on the ultraviolet completion and coupling structure, the EFT interpretation provides a model-independent estimate of the accessible scale.
\begin{equation}
\mathcal{L}_{W'}
\supset
g_q\,(\bar q \gamma^\mu P_L c)\,W'_\mu
+
g_\mu\,(\bar \mu \gamma^\mu P_L \nu)\,W'_\mu
+\text{h.c.}
\end{equation}
which generates the operator $\mathcal{O}_{V_L}$ at tree level. After
integrating out the heavy mediator, one obtains
\begin{equation}
C_{V_L}^{\mu}
=
\frac{g_q g_\mu\,v^2}{M_{W'}^2},
\label{eq:CVL_Wprime}
\end{equation}
with $v=246~\mathrm{GeV}$.

To make the impact of the baryonic observable explicit, we construct
$\Delta\chi^2$ profiles for both the meson-only and the combined
meson+baryon fits as functions of the single Wilson coefficient
$C_{V_L}^{\mu}$. These constraints are then mapped onto the ultraviolet
parameter space using Eq.~\eqref{eq:CVL_Wprime}, thereby embedding the
confidence-level structure of the EFT fit into the
$(M_{W'},\, g_q g_\mu)$ plane.

We further consider the difference
$\Delta\chi^2_{\mathrm{meson}} - \Delta\chi^2_{\mathrm{combined}}$ as a
diagnostic quantity to identify regions where the baryonic observable
provides additional constraining power. This quantity does not define a likelihood function, but instead provides a diagnostic measure of the local improvement in the constraint induced by the baryonic observable.

\begin{figure}[t]
	\centering
	\includegraphics[width=0.98\linewidth]{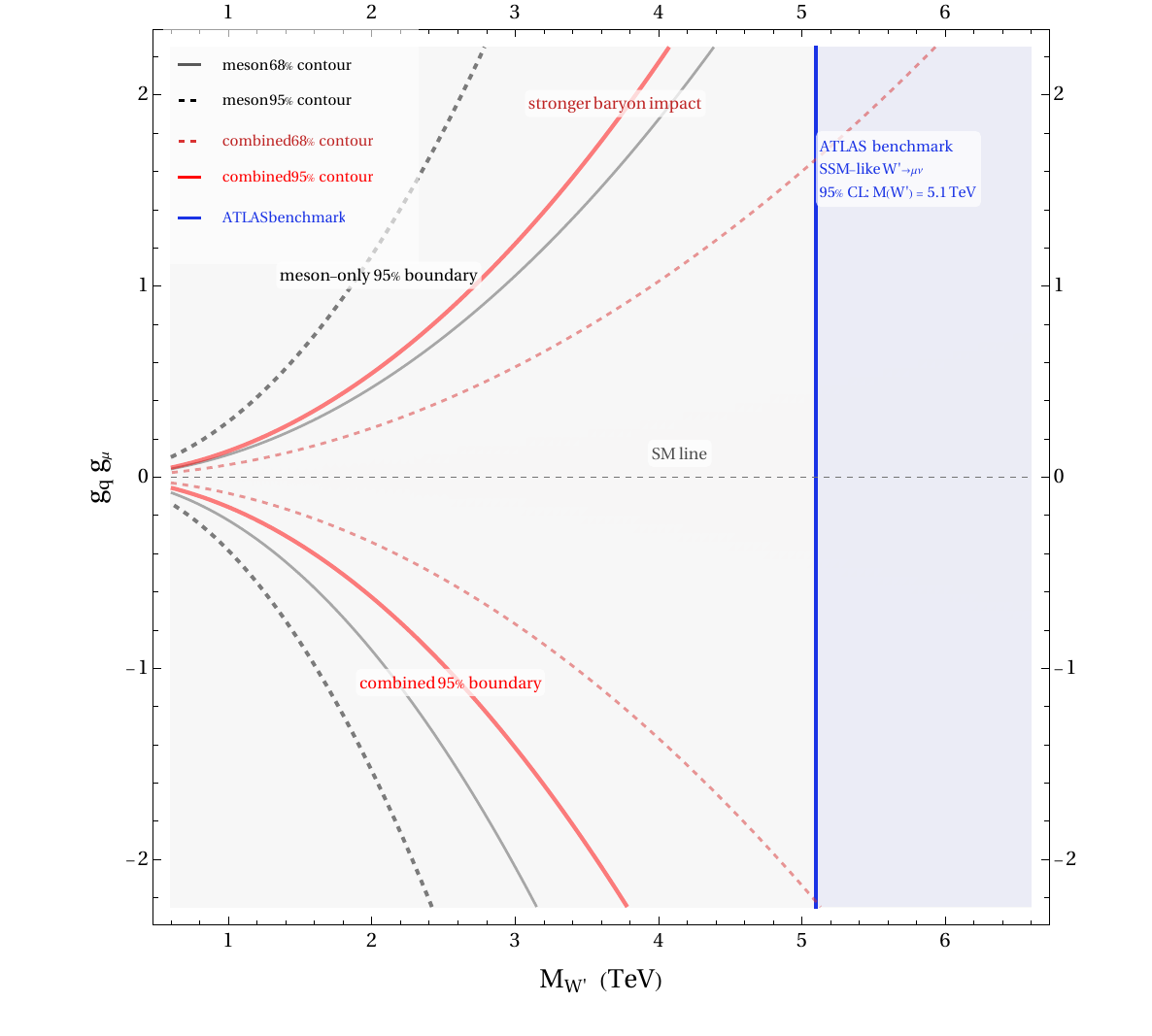}
	\caption{
		Ultraviolet mapping of the EFT constraint on $C_{V_L}^{\mu}$ in the
		$(M_{W'},\, g_q g_\mu)$ plane for a charged vector mediator with
		left-handed couplings. The contours correspond to constant
		$\Delta\chi^2$ derived from the meson-only fit (black) and from the
		combined meson+baryon fit (red), with solid and dashed lines indicating
		the $95\%$ and $68\%$ confidence levels, respectively.
		The heatmap shows the difference
		$\Delta\chi^2_{\text{meson}} - \Delta\chi^2_{\text{combined}}$,
		highlighting the regions where the inclusion of the projected baryonic
		observable $R_{\Xi_c}^{\mu e}$ provides additional constraining power.
		The meson input is based on the Belle II light-lepton universality
		measurement $R_{\text{mes}} = 1.007 \pm 0.021$,~\cite{BelleII:2023RXemu} while the baryonic contribution assumes a Standard Model prediction
		$R_{\Xi_c}^{\mu e} = 0.976$ Sec.~\ref{sec:semileptonic}. with a projected $1\%$ experimental precision. The vertical blue line indicates the ATLAS direct-search
		benchmark for a sequential standard-model-like
		$W' \to \mu\nu$ resonance,corresponding to an exclusion of $M_{W'} \simeq 5.1~\mathrm{TeV}$~\cite{ATLAS:2021wprime} for a sequential standard-model-like $W'$ boson The figure demonstrates that baryonic observables tighten existing
		bounds and reshape the allowed ultraviolet parameter space, providing
		complementary sensitivity to charged-current new physics.
	}
	\label{fig:WprimePlane}
\end{figure}

\paragraph{Two-dimensional parameter space.}
The EFT constraints can be mapped directly onto the
$(M_{W'},\,g_q g_\mu)$ plane, as shown in Fig.~\ref{fig:WprimePlane}. The meson-only fit defines an extended
allowed region, while the inclusion of the projected baryonic observable
$R_{\Xi_c}^{\mu e}$ leads to a significant contraction of this region.

Figure~\ref{fig:WprimePlane} shows that the inclusion of the projected
baryonic measurement leads to a non-trivial deformation of the allowed
ultraviolet parameter space. The effect is not limited to a uniform
reduction of the confidence region, but instead exhibits a pronounced
dependence on both the mediator mass and the effective coupling.

The heatmap reveals that the largest improvement occurs in regions of
moderate coupling and multi-TeV mediator masses, where the baryonic
observable provides sensitivity complementary to mesonic universality
tests. This behavior originates from the different parametric
dependence of the baryonic observable on the Wilson coefficient, which
enhances sensitivity away from the meson-dominated regime.

Importantly, the baryonic observable does not simply rescale the mesonic
constraint. Instead, it probes the same short-distance interaction with
a different parametric dependence, thereby reshaping the allowed
ultraviolet parameter space.

\paragraph{Envelope of constraints.}
To quantify the impact of the baryonic observable in a more direct and
model-independent manner, we extract the upper envelope of the allowed
coupling as a function of the mediator mass. Using the mapping
Eq.~\eqref{eq:CVL_Wprime}, the $95\%$ confidence intervals obtained in
the EFT fit can be translated into an upper bound on the effective
coupling combination $|g_q g_\mu|$ for each value of $M_{W'}$.
This representation isolates the maximal size of the new-physics
interaction compatible with low-energy data and allows for a direct
comparison of the constraining power of mesonic and baryonic
observables across the full parameter space.
\begin{equation}
\max |g_q g_\mu|(M_{W'}) \quad \text{at 95\% CL}.
\end{equation}
\begin{figure}[t]
	\centering
	\includegraphics[width=0.85\linewidth]{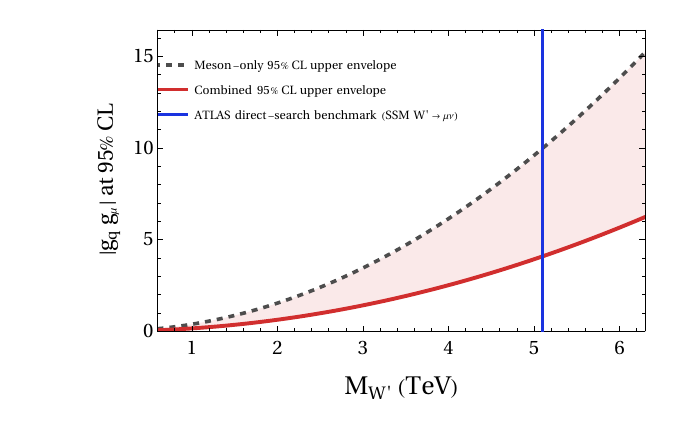}
	\caption{
		Upper envelope of the allowed effective coupling $|g_q g_\mu|$ as a
		function of the mediator mass $M_{W'}$ at $95\%$ confidence level.
		The dashed black curve corresponds to the constraint derived from the
		meson-only fit, based on the Belle II light-lepton universality
		measurement $R_{\text{mes}} = 1.007 \pm 0.021$ ~\cite{BelleII:2023RXemu}. The solid red curve includes the projected baryonic observable
		$R_{\Xi_c}^{\mu e}$, assuming a Standard Model value
		$R_{\Xi_c}^{\mu e} = 0.976$ with a $1\%$ experimental precision.
		The shaded region between the two curves quantifies the reduction of the
		allowed parameter space induced by the baryonic input. This reduction
		is most pronounced in the multi-TeV region, where the baryonic observable
		provides sensitivity complementary to mesonic universality tests.
		The vertical blue line indicates the ATLAS direct-search benchmark for
		a sequential standard-model-like $W' \to \mu\nu$ resonance, corresponding
		to $M_{W'} = 5.1~\mathrm{TeV}$ at $95\%$ CL. This collider bound is shown
		for reference only, as it depends on assumptions about the production
		cross section and decay branching fractions and does not directly map
		onto the $(M_{W'}, g_q g_\mu)$ parameter space.
	}
	\label{fig:WprimeEnvelope}
\end{figure}
Figure~\ref{fig:WprimeEnvelope} provides a quantitative measure of the
maximum allowed strength of the charged-current new-physics interaction
as a function of the mediator mass. The inclusion of the baryonic observable leads to a reduction of the allowed coupling over the full mass range, with a pronounced effect in
the multi-TeV region.
This behavior reflects the fact that the baryonic observable probes the
same short-distance interaction through a distinct hadronic structure,
leading to a different sensitivity to the underlying operator compared
to mesonic observables. As a result, it probes complementary combinations
of Wilson coefficients and kinematic configurations, thereby enhancing
sensitivity in regions where mesonic observables alone are less
constraining. In particular, for mediator masses in the multi-TeV range,
the allowed coupling strength is significantly reduced, demonstrating
that baryonic semileptonic decays provide competitive constraints on
ultraviolet completions.

The comparison with the ATLAS benchmark illustrates that low-energy
measurements can probe parameter regions that overlap with the reach of
direct searches, while remaining sensitive to different combinations of
couplings. This highlights the complementarity between flavor physics
and collider experiments in constraining charged-current new physics.
This representation makes explicit the tightening of the constraint
across the full mass range once baryonic information is included.
The shaded band between the two curves corresponds to the reduction
of the allowed parameter space due to the additional observable.

\paragraph{Comparison with collider searches.}
For reference, we indicate the ATLAS direct-search constraint for a
sequential standard-model-like $W' \to \mu\nu$ resonance, based on
$139~\mathrm{fb}^{-1}$ of proton--proton collisions at $\sqrt{s}=13~\mathrm{TeV}$,
corresponding to an exclusion of $M_{W'} \simeq 5.1~\mathrm{TeV}$ at
$95\%$ CL~\cite{ATLAS:2021wprime}. Comparable searches have also been
performed by CMS in the same final state, yielding limits of similar
order in the multi-TeV range~\cite{CMS:2022wqf}.

We emphasize that this collider bound is obtained from limits on the
production cross section times branching ratio, $\sigma(pp \to W') \times
\mathrm{BR}(W' \to \mu\nu)$, and therefore constrains a different
combination of parameters than the effective coupling $g_q g_\mu$
considered here. The comparison should thus be interpreted as a
qualitative benchmark rather than a direct exclusion in the
$(M_{W'}, g_q g_\mu)$ plane.

\paragraph{Physical implications.}
The results demonstrate that percent-level measurements of
$R_{\Xi_c}^{\mu e}$ probe charged-current new physics at energy scales
well into the multi-TeV regime. The envelope representation shows that
the maximal allowed strength of the underlying interaction is
systematically reduced once baryonic information is included.

This reduction is not a simple rescaling of the mesonic constraint, but
reflects the fact that baryonic observables probe independent hadronic
structures and kinematic regimes. Consequently, they provide
complementary sensitivity to the same short-distance operator, leading
to a nontrivial reshaping of the allowed ultraviolet parameter space.

These results establish semileptonic $\Xi_{cc}$ decays as an independent
probe of charged-current new physics, with sensitivity that is
comparable in scale to direct collider searches while probing different
combinations of couplings.
\section{Conclusion}
\label{sec:conclusion}
We have developed a precision framework for doubly charmed baryon decays
that combines nonleptonic null tests, semileptonic lepton-universality
ratios, and effective-field-theory interpretations. The main result is
that $\Xi_{cc}$ decays admit observables in which leading hadronic
normalization effects either cancel or are symmetry protected, allowing
QCD dynamics and short-distance charged-current interactions to be
tested in complementary ways.
In the nonleptonic sector, we constructed the null observable
$\Delta_{\pi K}$, which vanishes in the heavy-diquark factorization
limit. A nonzero value therefore isolates nonfactorizable QCD effects,
$SU(3)$ breaking, and subleading heavy-diquark corrections without
relying on absolute decay-rate normalizations. This makes
$\Delta_{\pi K}$ a controlled diagnostic of long-distance baryonic
dynamics.
In the semileptonic sector, we identified $R_{\Xi_c}^{\mu e}$ as a clean
light-lepton universality ratio. The leading baryonic normalization
cancels in this observable, leaving it primarily sensitive to phase
space and to the Lorentz structure of the charged-current interaction.
Universal left-handed vector deformations cancel in the ratio, whereas
lepton-nonuniversal vector interactions generate unsuppressed deviations.
Scalar contributions remain helicity suppressed in the light-lepton
regime. This produces a clear hierarchy of operator sensitivity and
makes $R_{\Xi_c}^{\mu e}$ a direct discriminator of effective
charged-current structures.
We further showed that baryonic and mesonic observables constrain the
same Wilson coefficient with complementary dependence. The combined
meson+baryon analysis tightens and reshapes the allowed parameter space
relative to meson-only fits, rather than simply improving a single
uncertainty. In the universal $C_{V_L}$ benchmark, the comparison of
light-lepton, semitauonic, and baryonic inputs illustrates that a single
universal rescaling of the Standard Model current is not sufficient to
describe all observables simultaneously. This points to either flavor
nonuniversality or a more general operator structure.
Mapping the EFT constraints onto a charged-vector benchmark shows that
percent-level measurements of $R_{\Xi_c}^{\mu e}$ probe multi-TeV
mediator scales. The resulting sensitivity is complementary to direct
collider searches because low-energy observables constrain different
coupling combinations and hadronic environments.
Overall, our results demonstrate that doubly charmed baryon decays
provide observables in which QCD dynamics and short-distance
charged-current interactions can be disentangled in a controlled way.
The combination of nonleptonic null tests and semileptonic
lepton-universality ratios enables a direct probe of operator structure
beyond the Standard Model. These observables supply nonredundant
constraints that complement mesonic measurements and offer a concrete
path for incorporating $\Xi_{cc}$ decays into future global flavor
analyses.

\end{document}